\documentclass[10pt,aps,prd,fleqn,superscriptaddress,notitlepage,nofootinbib,preprintnumbers,nobalancelastpage]{revtex4-1}
\pdfoutput=1
\usepackage{amsmath,amssymb,graphicx,xspace,subcaption,tikz}
\newcommand{\eps}{\varepsilon}
\newcommand{\Alaric}{{\sc Alaric}\xspace}
\newcommand{\Dire}{{\sc Dire}\xspace}
\newcommand{\Sherpa}{{\sc Sherpa}\xspace}
\usepackage{hyperref}

\hypersetup{hidelinks,
  pdfauthor={Florian Herren,Stefan Hoeche,Frank Krauss,Daniel Reichelt,Marek Schoenherr},
  pdftitle={A new approach to color-coherent parton evolution},
  pdfkeywords={Parton Shower,QCD Resummation}
}

\newcommand{\affFermilab}{Fermi National Accelerator Laboratory, Batavia, IL, 60510, USA}
\newcommand{\affIPPP}{Institute for Particle Physics Phenomenology, Durham University, Durham DH1 3LE, UK}

\begin{document}
\preprint{FERMILAB-PUB-22-556-T,  IPPP/22/58, MCNET-22-14}
\title{A new approach to color-\-coherent parton evolution}
\author{Florian~Herren}\affiliation{\affFermilab}
\author{Stefan~H{\"o}che}\affiliation{\affFermilab}
\author{Frank~Krauss}\affiliation{\affIPPP}
\author{Daniel~Reichelt}\affiliation{\affIPPP}
\author{Marek~Sch{\"o}nherr}\affiliation{\affIPPP}

\begin{abstract}
    We present a simple parton-shower model that replaces the explicit angular ordering
    of the coherent branching formalism with a differentially accurate simulation of 
    soft-gluon radiation by means of a non-trivial dependence of the splitting functions
    on azimuthal angles. We introduce a global kinematics mapping and provide an analytic proof
    that it satisfies the criteria for next-to leading logarithmic accuracy.
    In the new algorithm, initial and final state evolution are treated on the same footing.
    We provide an implementation for final-state evolution in the numerical code
    \Alaric and present a first comparison to experimental data.
\end{abstract}

\maketitle

\section{Introduction}
\label{sec:intro}
Parton showers are a cornerstone of computer simulations for high-energy 
collider physics~\cite{Buckley:2011ms,Campbell:2022qmc}. They implement 
the evolution of QCD from the hard scales to be probed by experiments, 
to the low scale of hadronization, where the transition of quasi-free partons 
(the quarks and gluons of perturbative QCD) to observable hadrons occurs. 
In this process, a number of additional partons are generated according to
evolution equations that are based on the factorization properties 
of QCD amplitudes in the soft and collinear limits.
The most commonly used parton showers can be thought of as numerical 
implementations of the DGLAP equations~\cite{Dokshitzer:1977sg,Gribov:1972ri,
  Lipatov:1974qm,Altarelli:1977zs}, but various other approaches 
  exist~\cite{Lonnblad:1992tz,Kharraziha:1997dn,Jung:2000hk}.

The first generation of parton shower programs~\cite{Webber:1983if,Bengtsson:1986gz,
Bengtsson:1986et,Marchesini:1987cf,Andersson:1989ki} was developed four decades ago.
Implementations differed in the way in which the ordering inherent to the 
evolution equations was realized in the simulation, and how the kinematics 
of the emissions were set up. Color coherence, manifesting itself through
angular ordering~\cite{Mueller:1981ex,Ermolaev:1981cm,Dokshitzer:1982fh,Dokshitzer:1982xr,
  Dokshitzer:1982ia,Bassetto:1982ma} became a guiding principle for the construction
  of parton showers~\cite{Webber:1986mc,Bengtsson:1986hr}.
Some of these parton showers were also improved using spin correlation 
algorithms~\cite{Collins:1987cp,Knowles:1987cu,Knowles:1988vs,Knowles:1988hu}.
Increasing precision requirements, especially in preparation for the
Large Hadron Collider (LHC), mandated more precise Monte-Carlo simulations. 
The matching of parton showers to next-to-leading order 
calculations~\cite{Frixione:2002ik,Nason:2004rx,Frixione:2007vw,
  Alioli:2010xd,Hoeche:2010pf,Hoeche:2011fd,Alwall:2014hca} and the merging
of calculations for varying jet multiplicity~\cite{Catani:2001cc,Lonnblad:2001iq,
  Krauss:2002up,Alwall:2007fs,Hoeche:2009rj,Lonnblad:2011xx,Gehrmann:2012yg,
  Hoeche:2012yf,Frederix:2012ps,Lonnblad:2012ix,Platzer:2012bs} became focus points
of event generator development. The correspondence between fixed-order infrared 
subtraction schemes and parton showers was identified as central to a correct
matching procedure, leading to the construction of algorithms with a 
dipole-local momentum mapping and ordering in transverse 
momentum~\cite{Nagy:2005aa,Nagy:2006kb,Schumann:2007mg,Giele:2007di,
Platzer:2009jq,Hoche:2015sya,Fischer:2016vfv,Cabouat:2017rzi}.

These newly developed algorithms were found to have significant drawbacks
in terms of their logarithmic accuracy~\cite{Dasgupta:2018nvj}. The resummation 
of observables at leading logarithmic (LL) accuracy is relatively straightforward 
to achieve using a parton-shower algorithm. The resummation at next-to-leading 
logarithmic (NLL) precision however poses a number of challenges. 
The first generic technique to quantify the logarithmic accuracy of parton showers 
was presented in~\cite{Dasgupta:2018nvj,Dasgupta:2020fwr} and consists of a set 
of fixed-order and all-order criteria, which can broadly be classified as tests 
related to kinematic recoil effects, and tests of color coherence. 
In the present manuscript, we will discuss only kinematic effects.
A discussion of sub-leading color effects can be found for example
in~\cite{Nagy:2012bt,Platzer:2012np,Nagy:2014mqa,Nagy:2015hwa,Platzer:2018pmd,
  Isaacson:2018zdi,Nagy:2019rwb,Nagy:2019pjp,Forshaw:2019ver,Hoche:2020pxj,
  DeAngelis:2020rvq,Holguin:2020joq,Hamilton:2020rcu}. 
One of the main results of~\cite{Dasgupta:2018nvj} was that the kinematics mapping
in the transition from an $n$-particle to an $n+1$-particle final state should not
alter the existing momentum configuration in a way that distorts the effects
of the pre-existing emissions on observables. 
This criterion is formulated such that only the effects relevant at NLL precision
can be extracted, by taking the limit $\alpha_s\to 0$ at fixed $\lambda=\alpha_s\ln v$,
where $v$ is the observable to be resummed. 
The algorithms in~\cite{Schumann:2007mg,Hoche:2015sya,Cabouat:2017rzi} do not satisfy
the criteria for NLL precision, because their momentum mappings can generate recoil
whose effect on existing emissions at commensurate scales does not vanish 
in the $\alpha_s\to 0$ limit. It is important to note that these failures to agree
with known NLL resummation are not related to the effects of momentum 
and probability conservation discussed in~\cite{Hoche:2017kst}. 
In order to remedy the problem with NLL accuracy, new kinematics mapping schemes 
were developed in~\cite{Bewick:2019rbu,Forshaw:2020wrq,Dasgupta:2020fwr,vanBeekveld:2022zhl}. 
The main difference of the new dipole schemes in~\cite{Dasgupta:2020fwr,vanBeekveld:2022zhl}
compared to existing algorithms is that recoil is assigned according to the rapidity
of the emission in the frame of the hard process, rather than the dipole frame,
and that initial-state radiation is treated such that the interpretation 
of the hard system is unchanged for subsequent emissions.

We will approach the same problem from a different perspective. 
Recalling that color--\-coherent parton evolution is a consequence of the angular 
dependence of the soft eikonal, we will reformulate the radiator functions
of~\cite{Webber:1986mc} using a partial fractioning approach similar to the 
identified particle subtraction scheme in~\cite{Catani:1996vz}.
In addition, we note that in dipole and antenna showers the anti-collinear direction
is inextricably linked to the direction of the color spectator. 
By lifting this restriction, we are able to construct an algorithm which allows 
the entire QCD multipole to absorb the recoil from parton branching, 
independent of the number of pre-existing emissions, and independent of their kinematics.
The price for such a generic scheme is a dependence of the parton shower 
splitting functions on the azimuthal angle between the decay plane 
and the plane defined by the emitting parton and its color spectator. 
Our new formulation presents a major extension of existing parton shower formalisms
in this regard, and it introduces the most generic form of a spin-averaged 
splitting function in four dimensions, with a dependence on all three phase-space variables
of the radiated parton. Based on previous analyses~\cite{Hoche:2017iem,Dulat:2018vuy},
it seems plausible that this scheme will considerably simplify the inclusion 
of higher-order corrections to the splitting kernels. We provide a first implementation 
of the new algorithm in the numerical code \Alaric\footnote{
  \Alaric is an acronym for A Logarithmically Accurate Resummation In C++},
which will be made available as part of the event generator 
\Sherpa~\cite{Gleisberg:2003xi,Gleisberg:2008ta,Sherpa:2019gpd}.

This manuscript is organized as follows: In Sec.~\ref{sec:coherence} we revisit 
the soft singularity structure of QCD amplitudes and introduce our new decomposition 
of the soft eikonal. In Sec.~\ref{sec:kinematics} we discuss the novel phase-space mapping
and the corresponding phase-space factorization. In Sec.~\ref{sec:algorithm} we detail 
how soft and collinear emissions are generated in a probabilistic picture. 
Section~\ref{sec:logarithms} is dedicated to the analytic proof of logarithmic accuracy,
and the numerical validation in the $\alpha_s\to 0$ limit. Section~\ref{sec:results} 
presents first numerical results for the process $e^+e^-\to\,$hadrons, 
and Sec.~\ref{sec:summary} contains an outlook.

\section{The matching of soft to collinear radiators}
\label{sec:coherence}
We start the discussion by recalling the singularity structure of $n$-parton
QCD amplitudes in the infrared limits.

If two partons, $i$ and $j$, become collinear, the squared amplitude factorizes as
\begin{equation}
    _{n}\langle1,\ldots,n|1,\ldots,n\rangle_{n}=
    \sum_{\lambda,\lambda'=\pm}
    \,_{n-1}\Big<1,\ldots,i\!\!\backslash(ij),\ldots,j\!\!\!\backslash,\ldots,n\Big|
    \frac{8\pi\alpha_s}{2p_ip_j}P^{\lambda\lambda'}_{(ij)i}(z)
    \Big|1,\ldots,i\!\!\backslash(ij),\ldots,j\!\!\!\backslash,\ldots,n\Big>_{n-1}\;,
\end{equation}
where the notation $i\!\!\backslash$ indicates that parton $i$ is removed 
from the original amplitude, and where $(ij)$ is the progenitor of partons $i$ and $j$.
The functions $P^{\lambda\lambda'}_{ab}(z)$ are the spin-dependent DGLAP splitting functions. 
They depend on the momentum fraction $z$ of parton $i$ with respect to the mother parton, $(ij)$,
and on the helicities $\lambda$~\cite{Dokshitzer:1977sg,Gribov:1972ri,Lipatov:1974qm,Altarelli:1977zs}.
In the collinear limit, the momentum fraction is equal to an energy or light-cone momentum fraction. 
In this manuscript we will consider only spin-averaged splitting functions; algorithms for 
spin-dependent evolution are discussed in~\cite{Collins:1987cp,Knowles:1987cu,Knowles:1988vs,
  Knowles:1988hu,Hamilton:2021dyz}.

In the limit that gluon $j$ becomes soft, the squared amplitude factorizes as~\cite{Bassetto:1984ik}
\begin{equation}
    _{n}\langle1,\ldots,n|1,\ldots,n\rangle_{n}=-8\pi\alpha_s\sum_{i,k\neq j}
    \,_{n-1}\big<1,\ldots,j\!\!\!\backslash,\ldots,n\big|{\bf T}_i{\bf T}_k\,w_{ik,j}
    \big|1,\ldots,j\!\!\!\backslash,\ldots,n\big>_{n-1}\;,
\end{equation}
where ${\bf T}_i$ and ${\bf T}_k$ are the color insertion operators defined in~\cite{Catani:1996vz}.
In the remainder of this section we will discuss the case of massless radiators only
and focus on the eikonal factor, $w_{ik,j}$, and how it can be rewritten in a suitable form
to match the spin-averaged splitting functions $P_{ab}(z)$ in the soft-collinear limit. 
Since our analysis concerns only the denominator of $w_{ik,j}$, it will apply to 
spin-correlated evolution as well. The eikonal factor is given by
\begin{equation}
    \label{eq:soft_eikonal_intro}
    w_{ik,j}=\frac{p_ip_k}{(p_ip_j)(p_jp_k)}\;,
\end{equation}
and it can be written in terms of (frame-dependent) energies and angles as
\begin{equation}
    \label{eq:soft_eikonal}
    w_{ik,j}=\frac{W_{ik,j}}{E_j^2}\;,
    \qquad\text{where}\qquad
    W_{ik,j}=\frac{1-\cos\theta_{ik}}{(1-\cos\theta_{ij})(1-\cos\theta_{jk})}\;,
\end{equation}
We note that Eq.~\eqref{eq:soft_eikonal} is symmetric in $i$ and $k$, and that it encapsulates
the complete soft singularity structure of the hard matrix element~\cite{Bassetto:1984ik}.
If we were to implement Eq.~\eqref{eq:soft_eikonal} for each of the radiators $i$ and $k$ 
in the collinear limit, we would therefore double-count the most singular component of the 
emission probability~\cite{Ellis:1991qj}. This is known as the soft double-counting problem, 
which can be solved by following the technique of \cite{Webber:1986mc}. 
In this approach, $W_{ik,j}$ is written as a sum of two terms, which are enhanced only 
in either the $ij$- or $kj$-collinear limit:
\begin{equation}
  \label{eq:additive_soft_matching}
  W_{ik,j}=\tilde{W}_{ik,j}^i+\tilde{W}_{ki,j}^k\;,
  \qquad\text{where}\qquad
  \tilde{W}_{ik,j}^i=\frac{1}{2}\left(\frac{1-\cos\theta_{ik}}{(1-\cos\theta_{ij})(1-\cos\theta_{jk})}
  +\frac{1}{1-\cos\theta_{ij}}
  -\frac{1}{1-\cos\theta_{jk}}\right)\;.
\end{equation}
It is customary to define the $z$-axis to be aligned with the momentum $p_i$, such that we can write
$\cos\theta_{jk}$ in terms of polar angles, $\theta_j^{\,i}$, $\theta_k^{\,i}$ with respect
to the axis defined by $p_i$, and the azimuthal angle $\phi_{jk}^{\,i}$ in the same frame.
Note in particular that $\theta_l^{\,i}=\theta_{li}$, for any $l$. 
\begin{equation}
    \cos\theta_{jk}=\cos\theta_j^{\,i}\cos\theta_k^{\,i}
    +\sin\theta_j^{\,i}\sin\theta_k^{\,i}\cos\phi_{jk}^{\,i}\;.
\end{equation}
When performing the azimuthal averaging, we find the simple result~\cite{Webber:1986mc}
\begin{equation}
  \label{eq:avg_additive_soft_matching}
    \frac{1}{2\pi}\int_0^{2\pi}{\rm d}\phi_{jk}^i\tilde{W}_{ik,j}^i=
    \frac{\tilde{I}_{ik,j}^i}{1-\cos\theta_j^{\,i}}\;,
    \qquad\text{where}\qquad
    \tilde{I}_{ik,j}^i=\left\{\begin{array}{cc}
    1 &\quad\text{if}\quad\theta_j^{\,i}<\theta_k^{\,i}\\[2mm]
    0 &\quad\text{else}\end{array}\right.\;.
\end{equation}
The behavior of $\tilde{I}_{ik,j}^i$ as a function of the polar angles is known as angular ordering,
which means that the total probability for soft radiation averages to zero outside of a cone 
defined by the cusp angle $\theta_k^{\, i}$ of the radiating color dipole.
This is the origin of the coherent branching formalism and the basis for angular ordered parton showers. 
It is instructive to investigate this radiation pattern in more detail.
\begin{figure}[t]
    \begin{subfigure}[c]{0.33\textwidth}
         \hspace*{5mm}\includegraphics[scale=0.45]{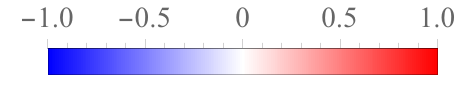}
         \includegraphics[scale=0.45]{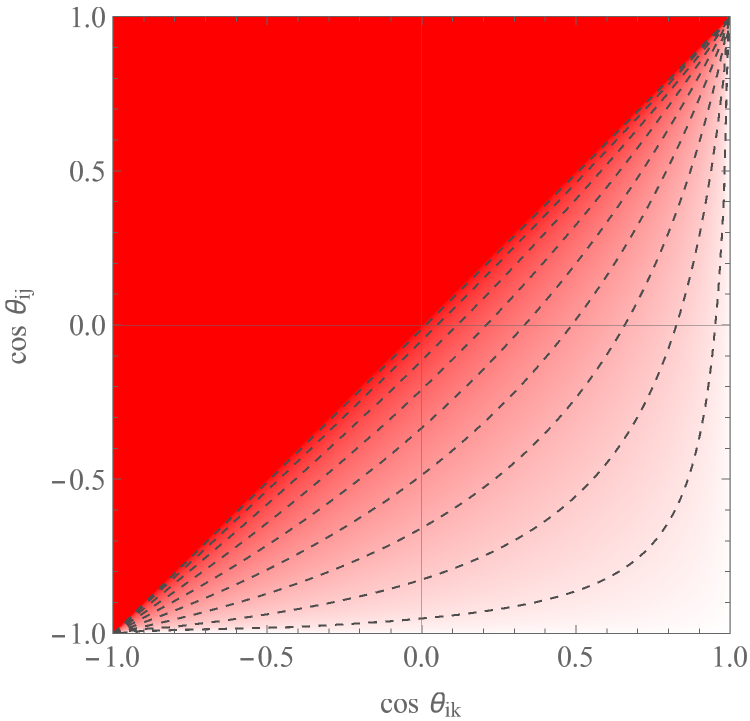}
         \caption{}
         \label{fig:phi_integral_additive_pos}
     \end{subfigure}\hfill
    \begin{subfigure}[c]{0.33\textwidth}
         \hspace*{5mm}\includegraphics[scale=0.45]{fig/heatscale.pdf}
         \includegraphics[scale=0.45]{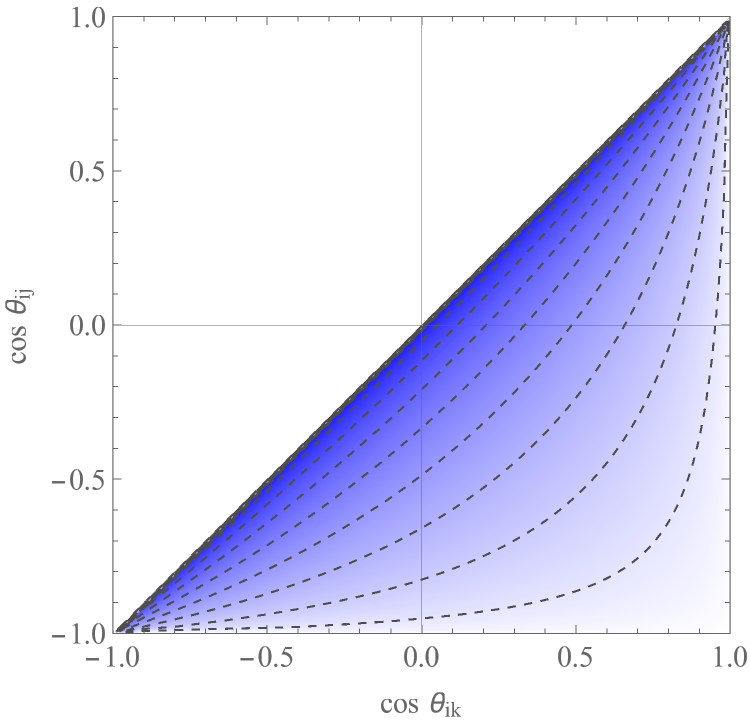}
         \caption{}
         \label{fig:phi_integral_additive_neg}
     \end{subfigure}\hfill
    \begin{subfigure}[c]{0.33\textwidth}
         \hspace*{5mm}\includegraphics[scale=0.45]{fig/heatscale.pdf}
         \includegraphics[scale=0.45]{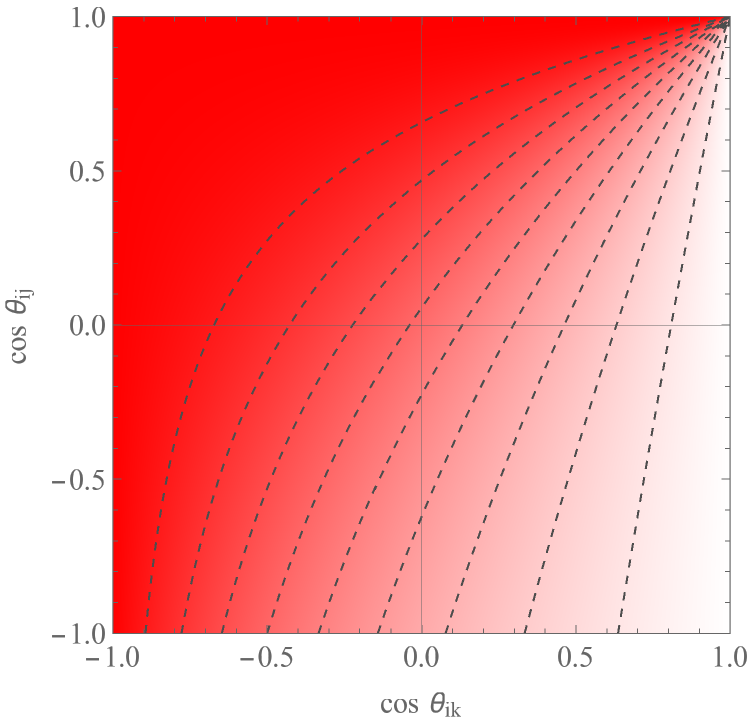}
         \caption{}
         \label{fig:phi_integral_partfrac}
     \end{subfigure}\hfill
    \caption{Azimuthally integrated radiator functions. 
    Figures~\subref{fig:phi_integral_additive_pos}
    and~\subref{fig:phi_integral_additive_neg} show the 
    positive and negative contributions to $\tilde{I}_{ik,j}^i$
    arising from the additive matching in 
    Eq.~\eqref{eq:additive_soft_matching},
    Fig.~\subref{fig:phi_integral_partfrac} displays
    $\bar{I}_{ik,j}^i$ from the multiplicative matching in
    Eq.~\eqref{eq:partfrac_soft_matching}.}
    \label{fig:phi_integrals}
\end{figure}
Figures~\ref{fig:phi_integral_additive_pos} and~\ref{fig:phi_integral_additive_neg} 
display the positive and negative contribution to the azimuthal integral,
normalized to $2\pi$, as a function of the polar angles.
The partial radiator function $\tilde{W}_{ik,j}^i$ has a root at
\begin{equation}
    \cos\phi_{jk}^{\,i(0)}=-\sqrt{\frac{1+\cos\theta_j^{\,i}}{1-\cos\theta_j^{\,i}}
    \,\frac{1-\cos\theta_k^{\,i}}{1+\cos\theta_k^{\,i}}}
\end{equation}
which falls inside the integration domain if $\theta_j^{\,i}>\theta_k^{\,i}$. 
In this case, the negative contribution to the azimuthal integral is equal in magnitude
to the positive contribution, such that the average radiation probability vanishes identically.
However, there is a strong modulation of this probability as a function of the azimuthal angle. 
If this modulation is not included in a parton-shower simulation, 
wide-angle soft radiation effects will only be captured correctly for observables
that are sufficiently insensitive to the precise distribution of radiation in phase space.

A naive attempt to solving this problem would be to include the full azimuthal dependence 
of the radiator function in the Monte-Carlo simulation. Such an approach is bound to fail,
because in the region $\theta_j^{\,i}>\theta_k^{\,i}$ one would need to sample the same amount
of negative and positive weighted Monte-Carlo events, leading to an efficiency of exactly zero. 
We therefore adopt a different strategy, pioneered in~\cite{Catani:1996vz}, where the radiator
function is partial fractioned such that it maintains strict positivity
\begin{equation}
    \label{eq:partfrac_soft_matching}
    W_{ik,j}=\bar{W}_{ik,j}^i+\bar{W}_{ki,j}^k\;,
    \qquad\text{where}\qquad
    \bar{W}_{ik,j}^i=\frac{1-\cos\theta_{ik}}{
      (1-\cos\theta_{ij})(2-\cos\theta_{ij}-\cos\theta_{jk})}\;.
\end{equation}
Azimuthal averaging again leads to Eq.~\eqref{eq:avg_additive_soft_matching}, 
except that $\tilde{I}_{ik,j}^i$ is replaced by
\begin{equation}
    \label{eq:avg_partfrac_soft_matching}
    \bar{I}_{ik,j}^i=
    \frac{1}{\sqrt{(\bar{A}_{ij,k}^{\,i})^2-(\bar{B}_{ij,k}^{\,i})^2}}\;,
\end{equation}
where
\begin{equation}
    \bar{A}_{ij,k}^{\,i}=\frac{2-\cos\theta_j^{\,i}(1+\cos\theta_k^{\,i})
    }{1-\cos\theta_k^{\,i}}\;\qquad\text{and}\qquad
    \bar{B}_{ij,k}^{\,i}=\frac{\sqrt{(1-\cos^2\theta_j^{\,i})
    (1-\cos^2\theta_k^{\,i})}}{1-\cos\theta_k^{\,i}}\;.
\end{equation}
This function is shown in Fig.~\ref{fig:phi_integral_partfrac}.
As required, it approaches unity in the limit $\theta_j^{\,i}\to 0$, 
independent of the value of $\theta_k^{\,i}$, and also for the special case
of a back-to-back configuration, $\theta_k^{\,i}\to \pi$.
While the Monte-Carlo efficiency of an algorithm using this technology 
will be reduced compared to plain angular ordered evolution, the obvious benefit
is that Eq.~\eqref{eq:partfrac_soft_matching} allows to capture all angular 
correlations associated with the spin-summed soft eikonal, Eq.~\eqref{eq:soft_eikonal}. 
In contrast, traditional angular ordered evolution, which is based 
on Eq.~\eqref{eq:avg_additive_soft_matching}, does not populate the complete 
emission phase space, necessitating intricate matrix-element corrections 
and creating complications in higher-order matching~\cite{Frixione:2002ik}.
We note again that the  energy $E_j$ in Eq.~\eqref{eq:soft_eikonal} is frame dependent.
This effect will be discussed in more detail in Sec.~\ref{sec:soft_algorithm}.

In the limit where partons $i$ and $j$ are collinear, 
we can write the eikonal factor in Eq.~\eqref{eq:soft_eikonal_intro} as
\begin{equation}
  w_{ik,j}\overset{i||j}{\longrightarrow}
  w_{ik,j}^{\rm(coll)}(z)=\frac{1}{2p_ip_j}\frac{2z}{1-z}\;,
  \qquad\text{where}\qquad
  z\overset{i||j}{\longrightarrow}\frac{E_i}{E_i+E_j}\;.
\end{equation}
This can be identified with the leading term (in $1-z$) of the
DGLAP splitting functions $P_{aa}(z)$, where~\footnote{
Note that in contrast to standard DGLAP notation, we separate the gluon
splitting function into two parts, associated with the soft singularities
at $z\to 0$ and $z\to1$.}
\begin{equation}\label{eq:dglap_splittings}
    \begin{split}
        P_{qq}(z)&=C_F\left(\frac{2z}{1-z}+(1-z)\right)\;,\\
        P_{gg}(z)&=C_A\left(\frac{2z}{1-z}+z(1-z)\right)\;,\\
        P_{gq}(z)&=T_R\left(1-2z(1-z)\right)\;.
    \end{split}
\end{equation}
To match the soft to the collinear splitting functions,
we therefore replace
\begin{equation}\label{eq:matching_soft_coll}
    \frac{1}{2p_ip_j}P_{(ij)i}(z)\to \frac{1}{2p_ip_j}P_{(ij)i}(z)
    +\delta_{(ij)i}\,{\bf T}_{i}^2
    \Bigg[\frac{\bar{W}_{ik,j}^i}{E_j^2}-w_{ik,j}^{\rm(coll)}(z)\Bigg]\;,
\end{equation}
where the two contributions to the gluon splitting function are treated 
as two different radiators~\cite{Hoche:2015sya}. This substitution introduces 
a dependence on a color spectator, $k$, whose momentum defines a direction 
independent of the direction of the collinear splitting. In general, 
this implies that splitting functions which were formerly dependent 
only on a momentum fraction along this direction, now acquire a dependence 
on the remaining two phase-space variables of the new parton. This is 
the most general form of a splitting kernel for spin-averaged parton evolution,
which we will use in the following.  In particular, the dependence on the
azimuthal angle allows to define the recoil momentum such that NLL precision is maintained
for any hard process, as discussed in more detail in Sec.~\ref{sec:logarithms}.

\section{Momentum mapping and phase-space factorization}
\label{sec:kinematics}
The mapping of Born momenta to a kinematic configuration after emission
of additional partons is a key component of any parton shower algorithm.
It is closely tied to the factorization of the Lorentz-invariant 
differential phase space element for a multi-parton configuration.
Suitable momentum mappings will preserve the key features of previously 
simulated radiation, while an unsuitable mapping could skew the QCD 
radiation pattern up to a point where it becomes not only theoretically
incorrect, but the differences become visible experimentally.
A prime, although academic, example for the latter problem is a collinear
unsafe mapping algorithm, in which the parton shower does not reflect 
the features of the collinear limit of the QCD matrix elements, 
Eq.~\eqref{eq:soft_eikonal_intro} and therefore introduces an error 
at leading logarithmic accuracy. A key requirement for the construction 
of any momentum mapping therefore is collinear safety, and all known
parton-shower algorithms satisfy this constraint.
An example for a problem which may only be seen in dedicated measurements
was identified in \cite{Dasgupta:2018nvj}. It originates in a modification
of existing soft momenta in subsequent emissions, that introduces an error
in the simulated QCD radiation pattern at next-to-leading logarithmic accuracy. 
In the following, we will construct a generic, collinear and NLL safe
momentum mapping for both final-state and initial-state radiation,
which is inspired by the identified--\-particle dipole subtraction 
algorithm in \cite{Catani:1996vz}. We will provide the analytic proof
of NLL safety in Sec.~\ref{sec:nll_proof} and sketch the additional steps
that are required to match the parton shower to NLO calculations 
in Appendix~\ref{sec:nlo matching}.

\begin{figure}[t]
\includegraphics[width=\textwidth]{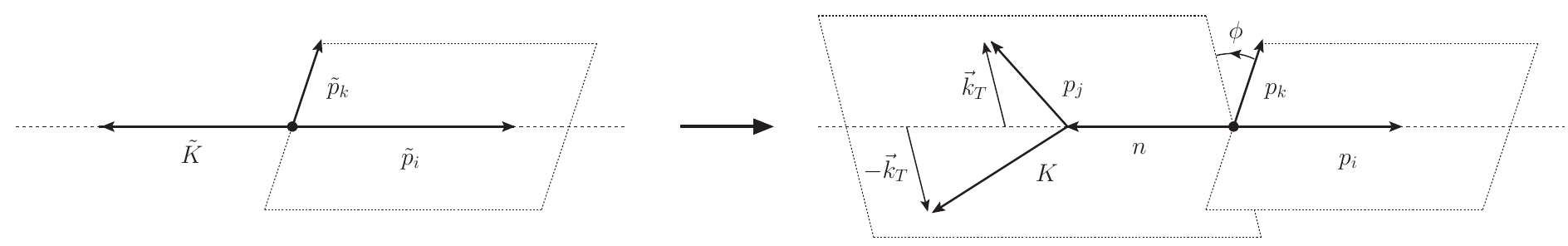}
\caption{Sketch of the momentum mapping for final-state evolution.
See the main text for details. Note that $p_k$ does not participate in the shift,
Eq.~\eqref{eq:def_n_pi}, and only acts as a reference for the azimuthal angle $\phi$.
\label{fig:kinematics_ff}}
\end{figure}
We begin by describing the logic underpinning our new kinematics mapping, 
$\{\tilde{p}_l\}\to\{p_l\}$. We identify the splitter momentum, $\tilde{p}_i$,
and define a recoil momentum, $\tilde{K}$, as the negative sum of all momenta 
in the radiating QCD multipole, {\it including} the momentum of the splitter 
(see also Appendix~\ref{sec:mc_details}).\footnote{This construction differs 
  from the traditional choice in parton and dipole showers, where the splitter
  and recoil partner are disjoint.} Together, the momenta $\tilde{K}$ and
$\tilde{p}_i$ define the reference frame of the splitting, as shown schematically 
in Fig.~\ref{fig:kinematics_ff} (left). The momentum of the color spectator,
$\tilde{p}_k$, defines an additional direction, which provides the reference
for the azimuthal angle, $\phi$. In the first step of the mapping, 
the emitter momentum is scaled by a factor $z$, and the emitted momentum,
$p_j$, is constructed with transverse momentum component $\vec{k}_T$ 
and suitable light-cone momenta. The color spectator remains unchanged,
$p_k=\tilde{p}_k$.  The recoil is absorbed by the overall multipole, 
such that after the emission we have $K\neq\tilde{K}$, while $K^2=\tilde{K}^2$. 
In particular, the multipole after the emission acquires a transverse momentum
with respect to $\tilde{K}$. This is shown schematically in 
Fig.~\ref{fig:kinematics_ff} (right). To compensate for both the transverse
and the longitudinal recoil, the overall multipole is boosted 
to its original frame of reference. This changes all momenta
and effectively distributes the recoil among them, generating changes
of the order of ${\rm k}_T/\sqrt{K^2}$, which vanish in the infrared limits.
We will make use of this fact in Sec.~\ref{sec:nll_proof}.

A collinear safe momentum mapping requires that for any two massless 
collinear partons, $i$ and $j$, the momenta behave as
\begin{equation}
  \label{eq:coll_mapping_exact_limit}
  p_i\overset{i||j}{\longrightarrow} z\,\tilde{p}_{i}\;,\qquad
  p_j\overset{i||j}{\longrightarrow} (1-z)\,\tilde{p}_{i}\;.
\end{equation}
In the exact limit, $\cos\theta_{ij}=0$, the splitting variable $z$
is uniquely defined and given by
\begin{equation}
    z=\frac{p_in}{(p_i+p_j)n}\;.
\end{equation}
where $n$ is an arbitrary auxiliary vector that satisfies $\tilde{p}_{i}n\neq 0$.
Note that $n$ can be either light-like, time-like or space-like, as long as
$\tilde{p}_{i}n\neq 0$. In order to construct a collinear-safe momentum mapping
for arbitrary values of the two-particle virtuality $p_ip_j$, we can simply use
the first part of Eq.~\eqref{eq:coll_mapping_exact_limit} away from this limit.
This implies in particular that $p_i$ retains its direction, and that all
angular radiator functions involving $p_i$ remain unchanged.

A second important constraint for the mapping is overall four-momentum conservation.
We satisfy this by defining a vector $\tilde{K}$ to be a combination of the momenta
$\{\tilde{p}_1,\ldots,\tilde{p}_{j-1},\tilde{p}_{j+1},\ldots,\tilde{p}_n\}$,
and by using the shift
\begin{equation}\label{eq:def_n_pi}
    p_i=z\,\tilde{p}_{i}\;,\qquad n=\tilde{K}+(1-z)\,\tilde{p}_{i}\;,
\end{equation}
which implies $p_i+n=\tilde{p}_{i}+\tilde{K}$.
The remaining task is to construct two new vectors, $K$ and $p_j$,
such that $K^2=\tilde{K}^2$, and such that $p_j$ satisfies
the collinear safety constraint, Eq.~\eqref{eq:coll_mapping_exact_limit}.
The momenta in $\tilde{K}$ are mapped to new momenta by a 
Lorentz transformation that is defined in terms of $\tilde{K}$ and $K$.
The simplest way to obtain the new momenta is by means
of a light-cone parametrization~\cite{Sudakov:1954sw}. 
With the help of the light-like vector
\begin{equation}
  \bar{n}=n-\frac{n^2}{2\tilde{p}_in}\,\tilde{p}_i
  =\tilde{K}-\kappa\,\tilde{p}_i\;,
  \qquad\text{where}\qquad
  \kappa=\frac{\tilde{K}^2}{2\tilde{p}_i\tilde{K}}\;.
\end{equation}
we can write
\begin{equation}\label{eq:def_pi_K}
    \begin{split}
        p_j&=v\,\bar{n}+\frac{1}{v}\frac{{\rm k}_\perp^2}{2\tilde{p}_i\tilde{K}}\,\tilde{p}_i-k_\perp\;,
        &\text{where}\qquad
        v&=\frac{p_ip_j}{p_i\tilde{K}}\\
        K&=(1-v)\,\bar{n}+\frac{1}{1-v}\frac{{\rm k}_\perp^2+\tilde{K}^2}{\,2\tilde{p}_i\tilde{K}}\,\tilde{p}_i+k_\perp\;.
    \end{split}
\end{equation}
Equation~\eqref{eq:def_pi_K} makes it manifest that $\tilde{K}$ absorbs 
the newly generated transverse and anti-collinear momentum when parton $(ij)$ 
is put off-shell, such that overall momentum conservation is satisfied.
This leads to the identity\footnote{In Eq.~\eqref{eq:def_kt2_identified},
  the variable $v$ takes the place of the splitting variable $z$
  in a standard collinear parametrization.}
\begin{equation}
  \label{eq:def_kt2_identified}
  {\rm k}_\perp^2=v(1-v)\,2p_jK-v^2K^2=
  v(1-v)(1-z)\,2\tilde{p}_i\tilde{K}-v^2\tilde{K}^2\;.
\end{equation}
Note that ${\rm k}_\perp^2$ is proportional to $v$ and therefore 
tends to zero in the collinear limit $\cos\theta_j^{\,i}\to 0$.
Inserting this relation into Eq.~\eqref{eq:def_pi_K} makes both 
collinear safety and overall four-momentum conservation 
of the kinematics mapping manifest.
\begin{equation}\label{eq:pi_K_using_kt2}
    \begin{split}
        p_j&=(1-z)\,\tilde{p}_i+v\big(\tilde{K}-(1-z+2\kappa)\,\tilde{p}_i\big)+k_\perp\;,\\
        K&=\hspace*{11mm}\tilde{K}-v\big(\tilde{K}-(1-z+2\kappa)\,\tilde{p}_i\big)-k_\perp\;.
    \end{split}
\end{equation}
In order to determine a reference direction for the azimuthal angle
$\phi=\arctan({\rm k}_y/{\rm k}_x)$, we note that the soft radiation pattern
of Eq.~\eqref{eq:partfrac_soft_matching} must be correctly generated. 
To achieve this we decompose the transverse momentum as
\begin{equation}\label{eq:likt}
  k_\perp^\mu={\rm k}_\perp\left(\cos\phi\, \frac{n_\perp^\mu}{|n_\perp|}
  +\sin\phi\, \frac{l_\perp^{\,\mu}}{|l_\perp|}\right)\;,
\end{equation}
where the reference axes $n_\perp$ and $l_\perp$ are given 
by the transverse projections\footnote{In kinematical configurations
  where $p_k^{\,\mu}$ is a linear combination of $p_i^{\,\mu}$ and
  $\bar{n}^{\,\mu}$, $n_\perp$ in the definition of Eq.~\eqref{eq:likt} vanishes. 
  It can then be computed using $n_\perp=\eps^{\mu j}_{\;\;\;\nu\rho}\, p_i^{\,\nu}\,\bar{n}^{\,\rho}$,
  where $j\in\{1,2,3\}$ may be any index that yields a nonzero result. 
  Note that in this case the matrix element cannot depend on the azimuthal angle.}
\begin{equation}
  n_\perp^\mu=p_k^\mu
  -\frac{p_k\bar{n}}{\tilde{p}_i\tilde{K}}\,\tilde{p}_i^\mu
  -\frac{p_k\tilde{p}_i}{\tilde{p}_i\tilde{K}}\,\bar{n}^\mu\;,
  \qquad\text{and}\qquad
  l_\perp^{\,\mu}=\eps^\mu_{\;\nu\rho\sigma}\,
  \tilde{p}_i^{\,\nu}\,\bar{n}^{\,\rho}\,n_\perp^\sigma\;.
\end{equation}
Because the differential emission phase-space element, 
Eq.~\eqref{eq:emission_phase_space}, is a Lorentz-invariant quantity,
the azimuthal angle $\phi$ is Lorentz invariant. It can be expressed as
\begin{equation}\label{eq:tc_ff_ps_dd_cosphi}
  \phi=\arccos\frac{(p_ip_j)(p_k{\bar{n}})+(p_ip_k)(p_j{\bar{n}})-(p_i{\bar{n}})(p_jp_k)}{
    \sqrt{2(p_ip_j)(p_j{\bar{n}})}\sqrt{2(p_ip_k)(p_k{\bar{n}})}}\;.
\end{equation}
This allows us to write the emission phase space in a frame-independent way.
After the momenta $p_i$, $p_j$ and $K$ are constructed, the momenta $\{p_l\}$ 
used to define $\tilde{K}$ are subjected to a Lorentz transformation, 
which can be written as~\cite{Catani:1996vz}
\begin{equation}\label{eq:lorentz_trafo_cs}
  p_l^\mu\to\Lambda^\mu_{\;\nu}(\tilde{K},K) \,p_l^\nu\;,  
  \qquad\text{where}\qquad
  \Lambda^\mu_{\;\nu}(\tilde{K},K)=g^\mu_{\;\nu}-\frac{2(K+\tilde{K})^\mu(K+\tilde{K})_\nu}{(K+\tilde{K})^2}
  +\frac{2K^\mu \tilde{K}_\nu}{\tilde{K}^2}\;.
\end{equation}
If needed, the event is restored to the lab frame, as described in App.~\ref{sec:mc_details}.
We list the precise algorithm for the construction of final-state splittings
in Sec.~\ref{sec:fs_ps}, and give the algorithm for the construction 
of initial-state splittings in Sec.~\ref{sec:is_ps}. 
The initial-state kinematics are obtained by the simple replacement 
$z\to x=1/z$, as indicated by crossing relations.

The last remaining task is to determine the differential emission 
phase space element. We will outline how to do this for pure 
final-state evolution, where the recoil partner $\tilde{K}$ corresponds
to the sum of the initial state momenta. This covers the important case
of the decay of a color-neutral, massive particle, such as a $Z$-boson at LEP.
Since $K$, $\tilde{K}$ and $n$ are treated as outgoing, they appear with 
negative energy component, which requires the explicit compensation of a number
of minus signs that do not appear in the case where $\tilde{K}$ corresponds
to a part of the final state. The general discussion can be found in 
App.~\ref{sec:ps_factorization}.

The differential phase-space element for $(n-1)$ hard momenta, 
$\{p_1,\ldots,p_{j-1},p_{j+1},\ldots,p_n\}$, 
and one soft momentum, $p_j$, is defined as
\begin{equation}\label{eq:final_phase_space}
  \begin{split}
    {\rm d}\Phi_{n}(p_a,p_b;p_1,\ldots,p_n)
    =&\;\left[\prod_{i=1}^{n}\frac{1}{(2\pi)^3}
    \frac{{\rm d}^3p_i}{2p_i^0}\right]
    (2\pi)^4\delta^{(4)}(p_a+p_b-\sum p_i)\;.
  \end{split}
\end{equation}
It can be written in terms of the differential phase-space element for the momenta
$\{\tilde{p}_1,\ldots,\tilde{p}_{j-1},\tilde{p}_{j+1},\ldots,\tilde{p}_n\}$ 
before the mapping
\begin{equation}\label{eq:born_phase_space}
  \begin{split}
    {\rm d}\Phi_{n-1}(\tilde{p}_a,\tilde{p}_b;\tilde{p}_1,\ldots,\tilde{p}_{j-1},\tilde{p}_{j+1},\ldots,\tilde{p}_n)
    =&\;\Bigg[\prod_{\substack{i=1\\i\neq j}}^{n}\frac{1}{(2\pi)^3}
    \frac{{\rm d}^3\tilde{p}_i}{2\tilde{p}_i^0}\Bigg]
    (2\pi)^4\delta^{(4)}(\tilde{p}_a+\tilde{p}_b-\sum_{i\neq j}\tilde{p}_i)\;,
  \end{split}
\end{equation}
and the ratio of differential phase-space elements after and before the mapping
\begin{equation}\label{eq:emission_phase_space}
  {\rm d}\Phi_{+1}(\tilde{p}_a,\tilde{p}_b;\tilde{p}_1,\ldots,\tilde{p}_{j-1},\tilde{p}_{j+1},\ldots,\tilde{p}_n;p_j)=
    \frac{{\rm d}\Phi_{n}(p_a,p_b;p_1,\ldots,p_n)}{
    {\rm d}\Phi_{n-1}(\tilde{p}_a,\tilde{p}_b;\tilde{p}_1,\ldots,\tilde{p}_{j-1},\tilde{p}_{j+1},\ldots,\tilde{p}_n)}\;.
\end{equation}
Eq.~\eqref{eq:emission_phase_space} denotes the single-emission phase space.
It can be computed using the lowest possible multiplicity, i.e.\ $n=2$
We start from the factorization formula\footnote{Note that the $n$-particle
differential phase-space element does not depend on the initial-state momenta
individually, hence the notation ${\rm d}\Phi_n(p_a,p_b;\ldots)$ is equivalent
to ${\rm d}\Phi_n(p_a+p_b;\ldots)$.}
\begin{equation}\label{eq:emission_phase_space_2to3}
    {\rm d}\Phi_{3}(-K;p_i,p_j,Q)
    ={\rm d}\Phi_{2}(-K;p_j,-n)\,
    \frac{{\rm d}n^2}{2\pi}\,
    {\rm d}\Phi_{2}(-n;p_i,Q)\;,
\end{equation}
where $Q=\sum_{k\neq i,j}p_k$. The two-particle
phase space in the frame of a time-like momentum $P$ can be written as
\begin{equation}\label{eq:two_body_ps}
    {\rm d}\Phi_{2}(p_1+p_2;p_1,p_2)=
    \frac{1}{16\pi^2}\frac{\sqrt{(p_1P)^2-p_1^2P^2}^{\,3}}{((p_1P)(p_1p_2)-p_1^2(p_2P))P^2}\,
    {\rm d}\cos\theta_1^{(P)}{\rm d}\phi_1^{(P)}\;,
\end{equation}
We perform all transformations in the rest frame of $n$, where we have 
the simple relations
\begin{equation}\label{eq:energies_n_frame_is}
    E_i=z\,\frac{-\tilde{p}_i\tilde{K}}{\sqrt{n^2}}\;,\qquad
    E_j=(1-z)\,\frac{-\tilde{p}_i\tilde{K}}{\sqrt{n^2}}\;,\qquad
    E_K=(1-z+2\kappa)\,\frac{-\tilde{p}_i\tilde{K}}{\sqrt{n^2}}\;,
    \quad\text{and}\quad
    n^2=-2\tilde{p}_i\tilde{K}\,(1-z+\kappa)\;.
\end{equation}
Using the following identity for the polar angle $\theta_j$ of the emission,
\begin{equation}\label{eq:ct_nframe_is}
    1-\cos\theta_j^{\,i}=2v\,\frac{1-z+\kappa}{1-z}\;,
\end{equation}
we find the first two-particle decay phase space
in Eq.~\eqref{eq:emission_phase_space_2to3} to be
\begin{equation}
    {\rm d}\Phi_{2}(-K;p_j,n)=
    \frac{1}{16\pi^2}\,
    {\rm d}v\,{\rm d}\phi_j^{(n)}\;.
\end{equation}
Note that this implies that $\phi_j^{(n)}$ is a Lorentz invariant quantity,
which is in fact given by Eq.~\eqref{eq:tc_ff_ps_dd_cosphi}. We also have
\begin{equation}\label{eq:n2_trafo_fi}
    {\rm d}n^2=-2\tilde{p}_i\tilde{K}\,{\rm d}z\;.
\end{equation}
Finally, we rewrite the second two-particle decay phase space as
\begin{equation}\label{eq:ps_fact_fi_ndecay}
    {\rm d}\Phi_{2}(-n;p_i,Q)=
    \frac{1}{16\pi^2}\frac{z}{2(1-z+\kappa)}\,
    {\rm d}\cos\theta_i^{(n)}\,{\rm d}\phi_i^{(n)}\;.
\end{equation}
In order to obtain a factorization formula, this must be mapped
to the Born phase space, which is given by
${\rm d}\Phi_{2}(-\tilde{K};\tilde{p}_i,\tilde{Q})$.
The angular integrals in Eq.~\eqref{eq:ps_fact_fi_ndecay}
are identical when working in the rest frame of the momentum $n$,
which leads to the relation
\begin{equation}
  {\rm d}\Phi_{2}(-n;p_i,Q)=z\,
    {\rm d}\Phi_{2}(-\tilde{K};\tilde{p}_i,\tilde{Q})\;.
\end{equation}
Combining all of the above, we find the single-emission 
phase space element
\begin{equation}
    {\rm d}\Phi_{+1}^{\rm(FI)}(-\tilde{K};\tilde{p}_1,\ldots,
    \tilde{p}_{j-1},\tilde{p}_{j+1},\ldots,\tilde{p}_n;p_j)
    =\frac{-2\tilde{p}_i\tilde{K}}{16\pi^2}\,
    {\rm d}v\,{\rm d}z\,z\,\frac{{\rm d}\phi}{2\pi}\;.
\end{equation}
We derive the analogous factorization formulae for recoilers in the final state 
and for initial-state emitters in App.~\ref{sec:ps_factorization}.

\section{Details of the algorithm}
\label{sec:algorithm}
This section introduces the details needed to implement our new parton-shower algorithm.
The procedure rests on the fact that the angular radiator function $(1-\cos\theta_j^{\,i})\bar{W}_{ik,j}^i$, 
with $\bar{W}_{ik,j}^i$ given in Eq.~\eqref{eq:partfrac_soft_matching}, has a fairly mild
dependence on the azimuthal angle. In particular, it is finite in the physical domain
$0<\theta_j^{\,i},\theta_k^{\,i}<\pi$. We can therefore generate the azimuthal angle using
a flat prior distribution, and work with standard algorithms for the remainder of the 
parton shower. In the following, we will assume some familiarity of the reader with these
algorithms. Details can be found in the many excellent reviews 
in the literature, for example~\cite{Ellis:1991qj,Sjostrand:2006za}.

\subsection{Soft evolution}
\label{sec:soft_algorithm}
We determine energies and angles in a global frame, which is defined by $n$. 
In the soft limit, $p_j\to 0$, this frame coincides with the event frame, defined by $K$.
The energies of particles $i$ and $j$ are given by Eq.~\eqref{eq:energies_n_frame_is}.
The polar angle $\theta_j^{\,i}$ of the emission is determined by
Eq.~\eqref{eq:ct_nframe_is}
\footnote{
  Note that for the first emission off a two-parton final state,
  $\kappa=-1$, such that $1-\cos\theta_j^{\,i}=2v\,z/(1-z)$, which is
  the same result as in the coherent branching formalism~\cite{Catani:1992ua}.}.
We define partial radiator functions, $\bar{w}_{ik,j}^i$, 
analogous to Eq.~\eqref{eq:partfrac_soft_matching},
such that $w_{ik,j}=\bar{w}_{ik,j}^i+\bar{w}_{ki,j}^k$.
This leads to
\begin{equation}
    \bar{w}_{ik,j}^i=\frac{\bar{W}_{ik,j}^i}{E_j^2}
    =\frac{\bar{W}_{ik,j}}{p_ip_j}\;,
    \qquad\text{where}\qquad 
    \bar{W}_{ik,j}=\frac{z}{1-z}\,(1-\cos\theta_j^{\,i})\,\bar{W}_{ik,j}^i\;.
\end{equation}
The function $\bar{W}_{ik,j}$ describes the frame-dependent azimuthal modulation
of the radiation pattern. We implement it in the numerically more convenient form
(see also Appendix~\ref{sec:nlo matching})
\begin{equation}\label{eq:frame_independent_radiator}
    \bar{W}_{ik,j}=\frac{l_{ik}p_i}{l_{ik}p_j}\;,
    \qquad\text{where}\qquad
    l_{ik}^\mu=\frac{p_i^\mu}{p_in}+\frac{p_k^\mu}{p_kn}\;.
\end{equation}
The function $(1-\cos\theta_j^{\,i})\bar{W}_{ik,j}^i$ assumes its maximum 
for $\phi_{jk}^{\,i}=0$. It is bounded from above by $2$. The eikonal part 
of the splitting function can therefore be overestimated by
\begin{equation}
  \label{eq:wijk_overestimate}
    \bar{w}_{ik,j}^i\le 2w_{ik,j}^{\rm(coll)}(z)=\frac{1}{2p_ip_j}\frac{4z}{1-z}\;.
\end{equation}
We define the evolution variable of the parton shower as
\begin{equation}\label{eq:def_t}
    t=2E_j^2\,(1-\cos\theta_j^{\;i})
    =v\,(1-z)\,2\tilde{p}_i\tilde{K}\;.
\end{equation}
Note that $1-\cos\theta_j^{\,i}=2\sin^2(\theta_j^{\,i}/2)$, such that $t$
corresponds to a transverse momentum. In the generalized rescaling limit 
$\rho\to 0$ of~\cite{Banfi:2004yd} (see Eq.~\eqref{eq:gen_rescaling_caesar}
and Sec.~\ref{sec:nll_proof} for details), it can be identified with the transverse momentum
squared in the Lund plane, hence our parton shower algorithm corresponds
to the case $\beta_{\rm PS}=0$ in~\cite{Dasgupta:2020fwr}.
The kinematical variable $v$ is given as a function of $t$ by
\begin{equation}
    v=\frac{\tau}{1-z}\;,
    \qquad\text{where}\qquad
    \tau=\frac{t}{2\tilde{p}_i\tilde{K}}\;.
\end{equation}
There is no Jacobian factor for the transformation $\ln v\to\ln t$.
The differential branching probability for soft radiation
is eventually given by the manifestly Lorentz invariant expression
\begin{equation}
    {\rm d}P_{ik,j}^{i\,\rm(soft)}(t,z,\phi)
    ={\rm d}\Phi_{+1}(\{\tilde{p}\},p_j)\,
    8\pi\alpha_s\,C_i\,\bar{w}_{ik,j}^i
    ={\rm d}t\,{\rm d}z\,\frac{{\rm d}\phi}{2\pi}\,
    \frac{\alpha_s}{2\pi\,t}\,2C_i\,\bar{W}_{ik,j}\;.
\end{equation}
For any $|\tau|>\tau_0=t_0/2\tilde{p}_i\tilde{K}$, with $t_0$ 
the infrared cutoff of the parton shower, Eq.~\eqref{eq:def_t}
defines a boundary on $z$ that is given by $z_+=1/(1+\tau_0)$.
This regularizes the integral of the overestimate of the splitting function 
in Eq.~\eqref{eq:wijk_overestimate}. We also introduce a lower bound on $z$,
given by $z_-^2=t_0/\tilde{K}^2$, to render the upper bound of the $t$ 
integration finite. This is analogous to the determination of the upper 
photon energy bound in~\cite{Schonherr:2008av}. The splitting variable $z$ 
can therefore be generated using standard Monte-Carlo techniques.

\subsection{Collinear evolution}
\label{sec:collinear_algorithm}
We are now left with the task to define the parton-shower algorithm to resum 
purely collinear logarithms. The corresponding splitting functions can be
derived by subtracting the collinear limit of the soft eikonal factor, 
Eq.~\eqref{eq:soft_eikonal_intro}, from the leading-order DGLAP 
splitting functions, Eq.~\eqref{eq:dglap_splittings}.
The differential branching probability for collinear radiation is then given by
(see Eq.~\eqref{eq:matching_soft_coll})
\begin{equation}
  \begin{split}
    {\rm d}P_{ik,j}^{i\,\rm(coll)}(t,z,\phi)
    =&\;{\rm d}\Phi_{+1}(\{\tilde{p}\},p_j)\,
    \frac{\alpha_s}{2\pi}\left(\frac{P_{\tilde{\imath}i}(z)}{2p_ip_j}
      -\delta_{\tilde{\imath}i}\,2C_i\,w_{ik,j}^{\rm(coll)}(z)\right)\\
    =&\;{\rm d}t\,{\rm d}z\,\frac{{\rm d}\phi}{2\pi}\,
    \frac{\alpha_s}{2\pi\,t}\left(P_{\tilde{\imath}i}(z)
      -\delta_{\tilde{\imath}i}\,C_i\,\frac{2z}{1-z}\right)\\
    =&\;{\rm d}t\,{\rm d}z\,\frac{{\rm d}\phi}{2\pi}\,
    \frac{\alpha_s}{2\pi\,t}\,C_{\tilde{\imath}i}\;.
  \end{split}
\end{equation}
Here we have defined the purely collinear remainder functions
\begin{equation}
    \begin{split}
        C_{qq}&=C_F\left(1-z\right)\;,\\
        C_{gg}&=C_A\,z(1-z)\;,\\
        C_{gq}&=T_R\left(1-2z(1-z)\right)\;.
    \end{split}
\end{equation}
While we use the same ordering parameter as in soft evolution, Eq.~\eqref{eq:def_t},
an ordering in virtuality or other variables is possible without affecting
the logarithmic precision.

\section{Analysis of logarithmic structure}
\label{sec:logarithms}
In this section we will analyze the logarithmic structure of the new parton-shower 
algorithm. We first provide an analytic proof that the recoil effects from new emissions
on pre-existing ones vanish in the $\rho\to 0$ limit~\cite{Banfi:2004yd}. This limit
corresponds to a similarity transformation in the Lund plane such that all emissions
can be considered as soft or collinear. The technique was introduced to eliminate 
corrections from kinematic effects which would generate terms beyond NLL accuracy.
Parton showers that create non-vanishing recoil effects in this limit are not 
NLL accurate~\cite{Dasgupta:2018nvj}.  Here we focus solely on the question 
whether the generalized scaling of emissions introduced in~\cite{Banfi:2004yd}
is maintained in our parton shower when additional splittings are generated
at lower or commensurate scales. In addition, we perform a numerical test 
of NLL accuracy, following the proposal in~\cite{Dasgupta:2020fwr}, 
which provides an additional strong check of our new algorithm.

\subsection{Recoil effects in the infrared limit}
\label{sec:nll_proof}
We will first show that the new kinematics mapping satisfies the fixed-order criteria
for NLL accuracy laid out in~\cite{Dasgupta:2018nvj,Dasgupta:2020fwr} to all orders.
Proofs for other parton-shower algorithms have been provided in numerical 
form~\cite{Dasgupta:2020fwr}, or based on approximations of the parton-shower 
branching probability, combined with analytical integration for specific 
observables~\cite{Nagy:2020dvz,Forshaw:2020wrq}. Here we will follow a different approach.
We describe the case of pure final-state evolution (for example in in $e^+e^-\to$hadrons),
similar arguments apply to initial-state evolution as well.

We follow~\cite{Banfi:2004yd} and denote the momenta of the hard partons as $p_1, \ldots, p_n$.
Additional soft emissions are denoted by $k$, and the observable we wish to compute by $v$.
In general, the observable will be a function of both the hard and the soft momenta,
$v=V(\{p\},\{k\})$, while in the soft approximation it reduces to a function of the 
soft momenta alone, $v=V(\{k\})$. In the rest frame of two hard legs, $i$ and $j$,
one may parametrize the momentum of a single emission as
\begin{equation}\label{eq:sudakov_decomposition}
  k=z_{i,j} p_i+z_{j,i} p_j+k_{T,ij}\;,
  \qquad\text{where}\qquad
  k_{T,ij}^2=2p_ip_j\,z_{i,j}\,z_{j,i}\;.
\end{equation}
The rapidity of the emission in this frame can be parametrized as 
$\eta_{ij}=1/2\ln(z_{i,j}/z_{j,i})$. The observable, computed as a function
of the momentum $k$, radiated collinear to the hard parton, $l$, can then be expressed as
\begin{equation}\label{eq:V_approx}
  V(k) = \left(\frac{k_{T,l}}{Q}\right)^a e^{-b_l\eta_l}\;,
\end{equation}
where, in the collinear limit, we have $k_{T,l}=k_{T,lj}$ and $\eta_l=\eta_{lj}$
for any $j\nparallel l$.

The cumulative cross section for an arbitrary observable, $v$, is defined as
\begin{equation}\label{eq:cumulative_xs}
  \Sigma\left(v\right) := \frac{1}{\sigma}\int^v \mathop{d\bar{v}}
  \frac{\mathop{d\sigma}}{\mathop{d\bar{v}}}\;.
\end{equation}
It is typically decomposed into a Sudakov factor, $e^{-R(v)}$, and a remainder function,
$\mathcal{F}(v)$, 
\begin{equation}
    \Sigma(v)=e^{-R(v)}\mathcal{F}(v)\;.
\end{equation}
The remainder function contains no leading logarithms, and the Sudakov radiator is a sum
over all partons in the hard process, $R=\sum_lR_l$. The function $\mathcal{F}(v)$
is extracted from the all-orders resummed result, Eq.~(2.34) of~\cite{Banfi:2004yd},
which reads
\begin{equation}
  \begin{split}
    \Sigma(v)=&\;\int{\rm d}^3k_1|M(k_1)|^2
    \exp\left\{-\int_{\eps v_1}{\rm d}^3k|M(k)|^2\right\}\\
    &\times\sum_{m=0}^{\infty} \frac{1}{m!}
    \bigg(\prod_{i=2}^{m+1}\int_{\eps v_1}^{v_1}{\rm d}^3k_i|M(k_i)|^2\bigg)
    \Theta\Big(v-V(\{p\},k_1,\ldots,k_n)\Big)\;.
  \end{split}
\end{equation}
A Taylor expansion in the virtual corrections up to first order in the derivative
of the Sudakov radiator, using a cutoff parameter, $\eps$, leads to
\begin{equation}
  \exp\left\{-\int_{\eps v_1}{\rm d}^3k|M(k)|^2\right\}
  =e^{-R(v)}\,e^{-R'\ln\frac{v}{\eps v_1}+\mathcal{O}(R^{\prime\prime})}\;,
  \quad\text{where}\quad
  R'=\frac{{\rm d}R}{{\rm d}\ln 1/v}\;.
\end{equation}
The function $M(k)$ is the single-emission matrix element in the infrared limit.
This leads to the convenient form (cf.\ Eq.~(2.37) in~\cite{Banfi:2004yd})
\begin{equation}
  \begin{split}
    \mathcal{F}(v)=&\;
    \int{\rm d}^3k_1|M(k_1)|^2\,e^{-R'\ln\frac{v}{\eps v_1}}
    \sum_{m=0}^{\infty} \frac{1}{m!}
    \bigg(\prod_{i=2}^{m+1}\int_{\eps v_1}^{v_1}{\rm d}^3k_i|M(k_i)|^2\bigg)
    \Theta\big(v-V(\{p\},k_1,\ldots,k_n)\big)\;.
  \end{split}
\end{equation}
Here, $v_1$ is the value of the observable in the leading (in $v$) emission,
and $k_1$ is the corresponding momentum. The expressions to the left of the sum
can be interpreted as the differential radiation probability for the first emission,
and the corresponding Sudakov suppression factor, assuming that further radiation
is resolved down to a scale of $\eps v_1$. The sum then implements the corresponding
real radiative corrections to all orders, while the $\Theta$ function accounts for the
constraint from the observable, $v$. This makes it clear that the $\mathcal{F}$ function
is due to multiple emission effects.

In order to cleanly extract the NLL expression for $\mathcal{F}(v)$, the limit $\eps\to 0$
must be taken, and the sum over emissions must be computed to all orders. This corresponds
to the limit $\alpha_s\to 0$ or $v\to 0$, while $\alpha_s\ln 1/v$ remains constant.
The case $v\to 0$ can be understood as the limit of infinite center-of-mass energy.
In this limit, kinematic edge effects can be neglected. However, it must be guaranteed
that the event topology in the limit remains the same as in a situation with finite $v$,
which implies that the observable must satisfy the recursive IRC safety conditions 
laid out in Sec.~2.2.3 of~\cite{Banfi:2004yd}. If it does, one will be able to take 
the limit and compute $\mathcal{F}(v)$ by performing a similarity transformation
in the Lund plane, which is given in terms of a scaling parameter, $\rho$,
by Eq.~(2.39) of~\cite{Banfi:2004yd}
\begin{equation}\label{eq:gen_rescaling_caesar}
  k_{t,l}\to k_{t,l}'=k_{t,l}\rho^{(1-\xi_l)/a+\xi_l/(a+b)}\;,
  \qquad
  \eta_l\to \eta_l'=\eta-\xi_l\frac{\ln\rho}{a+b}\;,
  \qquad\text{where}\qquad
  \xi=\frac{\eta}{\eta_{\rm max}}\;.
\end{equation}
This transformation is sketched in Fig.~3 of~\cite{Banfi:2004yd}. 
The aim of our proof is to show that the recoil arising from the inverse of the
Lorentz transformation in Eq.~\eqref{eq:lorentz_trafo_cs} does not lead to an 
appreciable alteration of the momenta of pre-existing emissions 
in the limit where the scaling parameter vanishes, $\rho\to 0$.\footnote{
  In the case of $e^+e^-\to$hadrons, Eq.~\eqref{eq:lorentz_trafo_cs} is applied to 
  move the initial-state momenta of the $e^+e^-$ collision to a new frame. Afterwards,
  Eq.~\eqref{eq:lt_fi} is applied to restore the complete event to the lab frame.
  As Eq.~\eqref{eq:lt_fi} in this case is the inverse of Eq.~\eqref{eq:lorentz_trafo_cs},
  this corresponds to applying the inverse of Eq.~\eqref{eq:lorentz_trafo_cs} to the 
  complete final state directly. All other scenarios can be treated in the same fashion
  for the purpose of this proof.}
In order to analyze the behavior of the Lorentz transformation, we switch back to
our original notation and use Eq.~\eqref{eq:pi_K_using_kt2} to split $K^\mu$ 
into its components along the recoil momentum, $\tilde{K}^\mu$, 
the emitter momentum, $\tilde{p}_i^\mu$, and the emission, $p_j^\mu$,
\begin{equation}\label{eq:lt_defining_momentum}
    \begin{split}
        K^\mu&=\tilde{K}^\mu-X^\mu\;,
        \qquad\text{where}\qquad
        X^\mu=p_j^\mu-(1-z)\,\tilde{p}_i^\mu\;.
    \end{split}
\end{equation}
The vector $X^\mu$ will tend to zero in both the soft and the collinear limit, because 
it has no component along the direction of the emitter momentum, $\tilde{p}_i$.
This implies in particular that for emissions off the original hard partons, 
$X^\mu$ will tend to zero, even in the hard collinear region, such that 
the Lorentz transformation vanishes. In terms of $\tilde{K}^\mu$ and $X^\mu$,
Eq.~\eqref{eq:lorentz_trafo_cs} takes the form
\begin{equation}\label{eq:lorentz_trafo_cs_ff}
  \Lambda^\mu_{\;\nu}(K,\tilde{K})=g^\mu_\nu+\tilde{K}^\mu A_\nu+X^\mu B_\nu\;,
\end{equation}
where
\begin{equation}
  A^\nu=2\bigg[\,\frac{(\tilde{K}-X)^\nu}{(\tilde{K}-X)^2}
    -\frac{(\tilde{K}-X/2)^\nu}{(\tilde{K}-X/2)^2}\,\bigg]\;,
  \qquad\text{and}\qquad
  B^\nu=\frac{(\tilde{K}-X/2)^\nu}{(\tilde{K}-X/2)^2}\;.
\end{equation}
Following Sec.~2.2.3 of~\cite{Banfi:2004yd}, we now analyze the behavior of 
this change under the generalized rescaling of all emissions, $p_l$, according to
Eq.~\eqref{eq:gen_rescaling_caesar}.
Note that the transverse momentum $k_t$ in this analysis is not the same 
as ${\rm k}_\perp$ in Eq.~\eqref{eq:pi_K_using_kt2}. It is instead given 
in terms of Lund plane coordinates, see Sec.~2 of~\cite{Banfi:2004yd} 
for details of these definitions. We can choose to use the initial momenta 
of the hard quark and anti-quark (which are not subject to the rescaling)
as reference directions to define the Lund plane transverse momentum and rapidity,
and work in their rest frame with the quark (antiquark) momentum pointing
along the positive (negative) $z$ direction. In this frame, the longitudinal
components of the momenta $p_l$ scale as $\tilde{p}_l^{0,3}\sim \rho^{(1-\xi_l)/a}$,
while the transverse components behave as $\tilde{p}_l^{1,2}\sim \rho^{(1-\xi_l)/a+\xi_l/(a+b)}$.

From Eq.~\eqref{eq:lt_defining_momentum} we deduce that all components
of $X^\mu$ scale as the soft momenta $\tilde{p}_l$ in
Eq.~\eqref{eq:gen_rescaling_caesar}, because the component of $p_j$
along the emitter momentum $\tilde{p}_i$ has been subtracted.
This is a very important feature of our kinematics mapping. 
We will now show that this mapping maintains the scaling properties,
Eq.~\eqref{eq:gen_rescaling_caesar}, of an arbitrary set of pre-existing
emissions in the $\rho\to 0$ limit.

First we take the $\rho\to 0$ limit of the coefficients in Eq.~\eqref{eq:lorentz_trafo_cs_ff}.
The leading contributions are given by
\begin{equation}
    A^\nu\overset{\rho\to 0}{\longrightarrow}
    2\,\frac{\tilde{K}X}{\tilde{K}^2}\,\frac{\tilde{K}^\nu}{\tilde{K}^2}
    -\frac{X^\nu}{\tilde{K}^2}\;,
    \qquad\text{and}\qquad
    B^\nu\overset{\rho\to 0}{\longrightarrow}\frac{\tilde{K}^\nu}{\tilde{K}^2}\;.
\end{equation}
The momentum shift of particle $l$ under the Lorentz transformation is then given by
\begin{equation}\label{eq:momentum_shift_under_lt}
    \Delta p_l^\mu=2\,\frac{\tilde{K}X}{\tilde{K}^2}\,
    \frac{\tilde{p}_l\tilde{K}}{\tilde{K}^2}\,\tilde{K}^\mu
    -\frac{\tilde{p}_lX}{\tilde{K}^2}\,\tilde{K}^\mu
    +\frac{\tilde{p}_l\tilde{K}}{\tilde{K}^2}\,X^\mu\;.
\end{equation}
For color singlet decay or production processes we can work in the multipole center-of-mass frame.
$\tilde{K}$ then only has an energy component, which is not rescaled as $\rho\to 0$.
Let us first assume that the emitter momentum, $\tilde{p}_i$, is one of the soft momenta.

The scaling of the scalar products in Eq.~\eqref{eq:momentum_shift_under_lt} is then given by
\footnote{Note that $\tilde{p}_lX$ has two contributions, one proportional to
  $\rho^{(2-(\xi_l+\max(\xi_i,\xi_j)))/a}$, and one proportional to
  $\rho^{(2-b/(a+b)(\xi_l+\max(\xi_i,\xi_j)))/a}$.
  The first one dominates in all cases, because $b/(a+b)<1$. 
  While $b$ can be negative, infrared and collinear safety
  requires $b > -a$, $a > 0$.}
\begin{equation}
    \begin{split}
    \tilde{p}_l\tilde{K} &\sim\rho^{(1-\xi_l)/a}\;,\\
    \tilde{p}_lX &\sim\rho^{(2-\xi_l-\max(\xi_i,\xi_j))/a}\;.
    \end{split}
\end{equation}
The denominators in $A^\nu$ and $B^\nu$ do not scale with $\rho$. 
With that we can derive the scaling of the change in each component of $p_l$ and compare it
to the scaling of the original components in $\tilde{p}_l$. 
\begin{equation}\label{eq:scaling_with_lt}
    \begin{split}
    \tilde{p}_l^0 \sim&\; \rho^{(1-\xi_l)/a} \qquad& \Delta p_l^0 
    \sim&\; \rho^{(1-\xi_l)/a} X^0 + \rho^{(2-\xi_l-\max(\xi_i,\xi_j))/a} \tilde{K}^0 + \rho^{(1-\xi_l)/a} X^0 
    \sim \rho^{(2-\xi_l-\max(\xi_i,\xi_j))/a}\;, \\
    \tilde{p}_l^3 \sim&\; \rho^{(1-\xi_l)/a} & \Delta p_l^3 
    \sim&\; \rho^{(1-\xi_l)/a} X^3 \sim \rho^{(2-\xi_l-\max(\xi_i,\xi_j))/a}\;,\\
    \tilde{p}_l^{1,2} \sim&\; \rho^{(1-b/(a+b)\xi_l)/a} & \Delta p_l^{1,2}
    \sim&\; \rho^{(1-\xi_l)/a} X^{1,2} \sim \rho^{(2-\xi_l-b/(a+b)\max(\xi_i,\xi_j))/a}\;.
    \end{split}
\end{equation}
The relative momentum shifts are
\begin{equation}
  \begin{split}
  \frac{\Delta p_l^{0,3}}{p_l^{0,3}} \sim&\; \rho^{(1-\max(\xi_i,\xi_j))/a}\;,\\
  \frac{\Delta p_l^{1,2}}{p_l^{1,2}} \sim&\; \rho^{(1-\xi_l-b/(a+b)(\max(\xi_i,\xi_j)-\xi_l))/a}
  < \rho^{(1-b/(a+b))(1-\xi_l)/a} \;.
  \end{split}
\end{equation}
If $\xi_l<1$ and $\max(\xi_i,\xi_j)<1$, these changes vanish in the $\rho\to 0$ limit.
The case of $\xi_l=1$ and/or $\max(\xi_i,\xi_j)=1$ corresponds to a phase-space region
of measure zero and does therefore not need to be considered.

In the case where $\tilde{p}_i$ is one of the hard momenta, the leading terms in
Eq.~\eqref{eq:lt_defining_momentum} cancel exactly, and the remaining components
of $X^\mu$ are transverse or anti-collinear, leading to a scaling with $\rho^{1/a}$
and $\rho^{2/a}$, respectively, in Eq.~\eqref{eq:scaling_with_lt}. This leads to the
same conclusions as the case $\xi_i=\xi_j=0$.

\begin{figure}[t]
  \centering
  \includegraphics[width=0.5\textwidth]{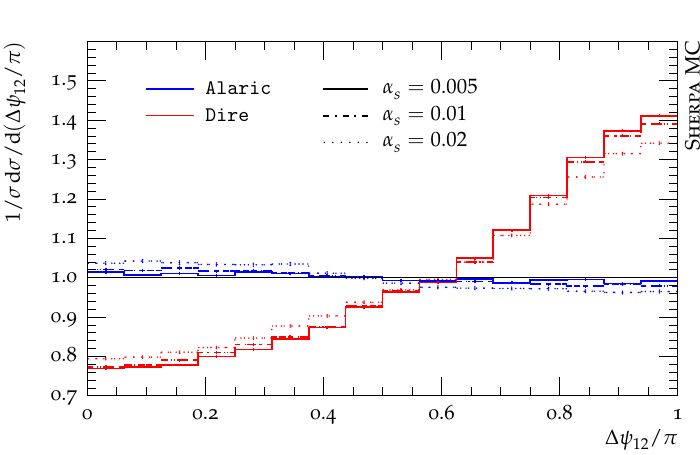}
  \caption{NLL test for $\Delta\psi_{12}$.\label{fig:nll_test_dpsi}}
\end{figure}
\subsection{Numerical tests of kinematics mapping}
\label{sec:nll_numerics}
\begin{figure}[p]
  \centering
  \begin{minipage}{5.875cm}
    \includegraphics[width=\textwidth]{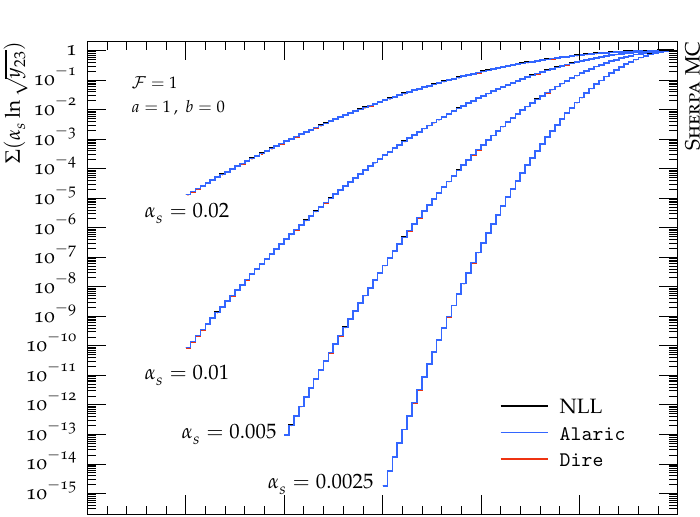}\\[-.8mm]
    \includegraphics[width=\textwidth]{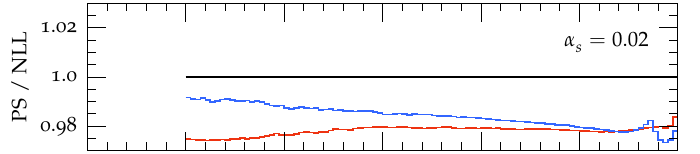}\\[-.8mm]
    \includegraphics[width=\textwidth]{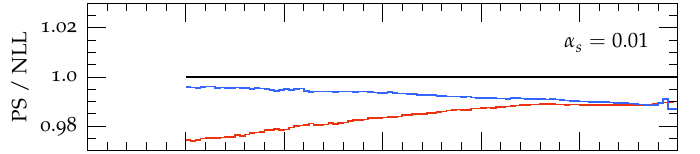}\\[-.8mm]
    \includegraphics[width=\textwidth]{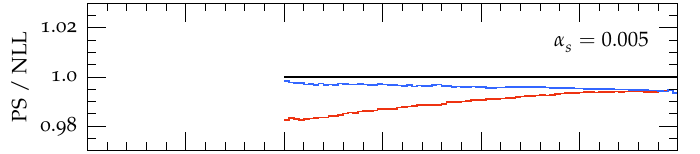}\\[-.8mm]
    \includegraphics[width=\textwidth]{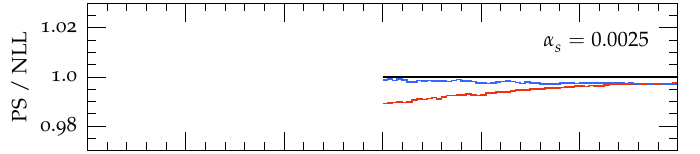}\\[-.8mm]
    \includegraphics[width=\textwidth]{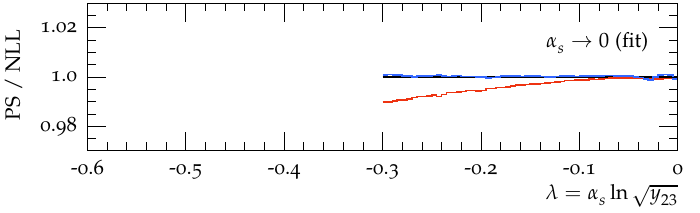}
  \end{minipage}\hfill
  \begin{minipage}{5.875cm}
    \includegraphics[width=\textwidth]{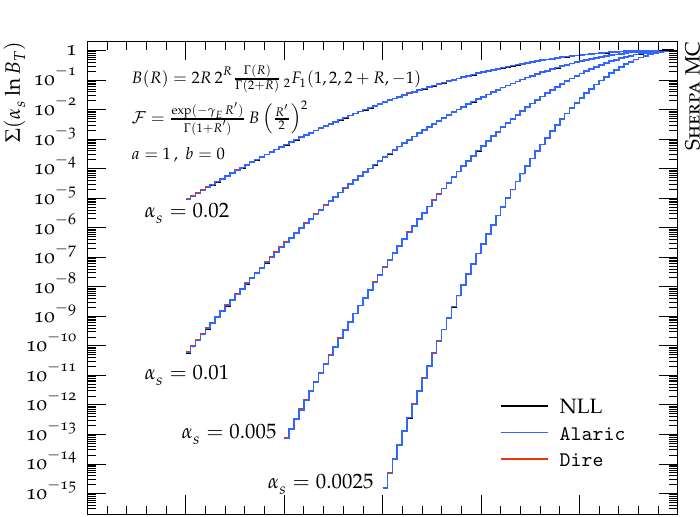}\\[-.8mm]
    \includegraphics[width=\textwidth]{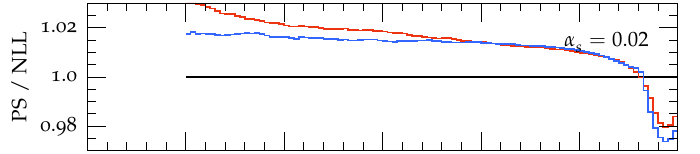}\\[-.8mm]
    \includegraphics[width=\textwidth]{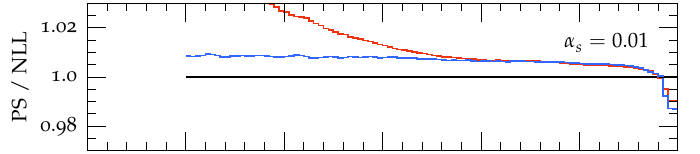}\\[-.8mm]
    \includegraphics[width=\textwidth]{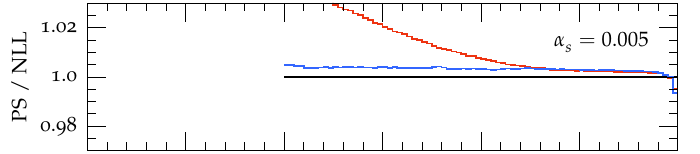}\\[-.8mm]
    \includegraphics[width=\textwidth]{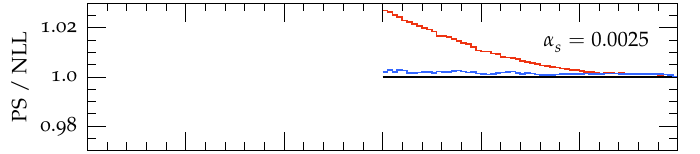}\\[-.8mm]
    \includegraphics[width=\textwidth]{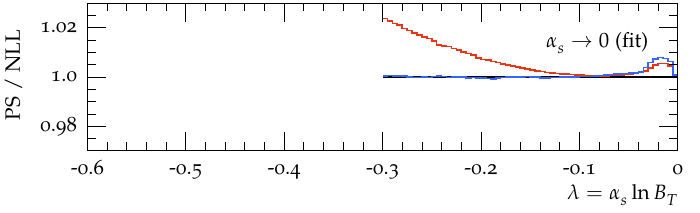}
  \end{minipage}\hfill
  \begin{minipage}{5.875cm}
    \includegraphics[width=\textwidth]{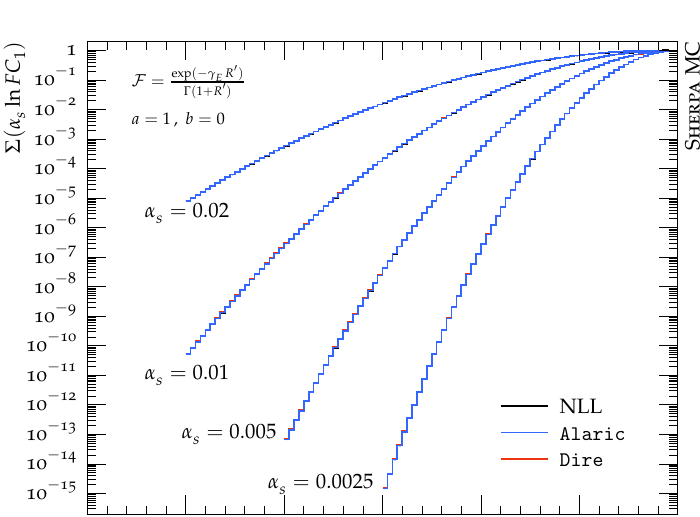}\\[-.8mm]
    \includegraphics[width=\textwidth]{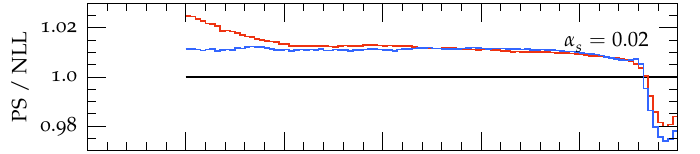}\\[-.8mm]
    \includegraphics[width=\textwidth]{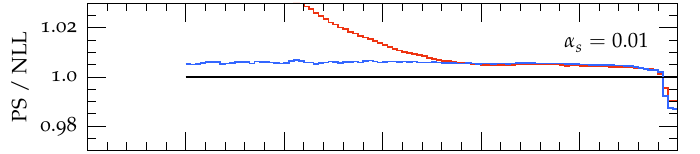}\\[-.8mm]
    \includegraphics[width=\textwidth]{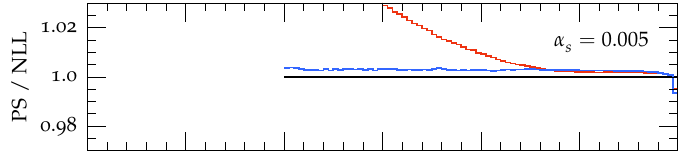}\\[-.8mm]
    \includegraphics[width=\textwidth]{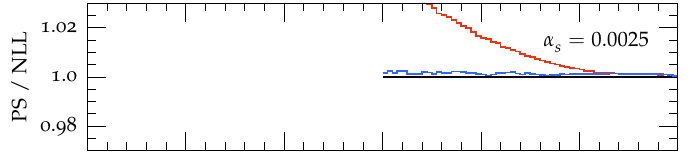}\\[-.8mm]
    \includegraphics[width=\textwidth]{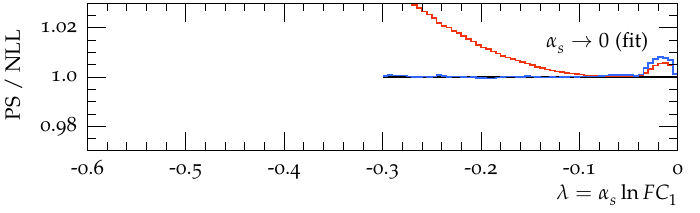}
  \end{minipage}\\
  \begin{minipage}{5.875cm}
    \includegraphics[width=\textwidth]{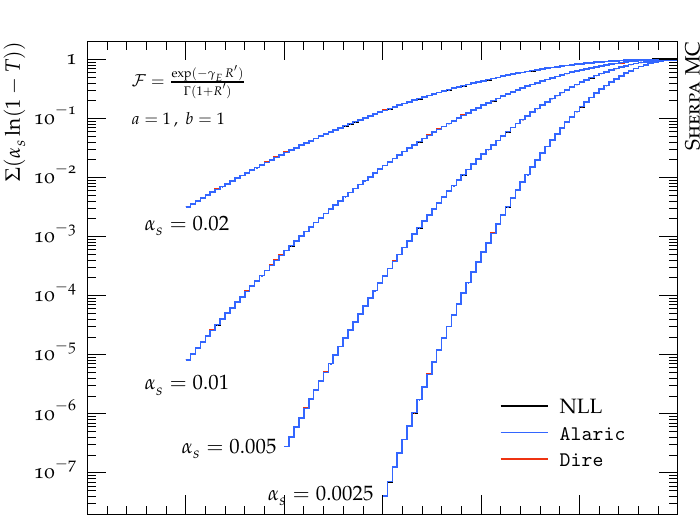}\\[-.8mm]
    \includegraphics[width=\textwidth]{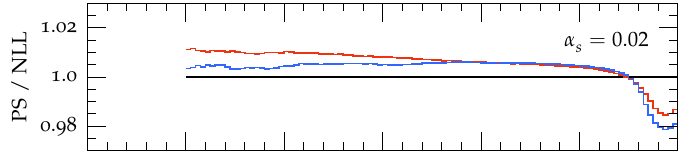}\\[-.8mm]
    \includegraphics[width=\textwidth]{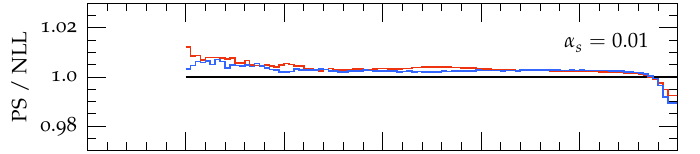}\\[-.8mm]
    \includegraphics[width=\textwidth]{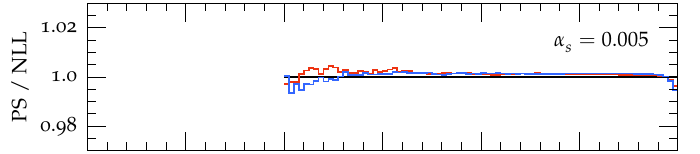}\\[-.8mm]
    \includegraphics[width=\textwidth]{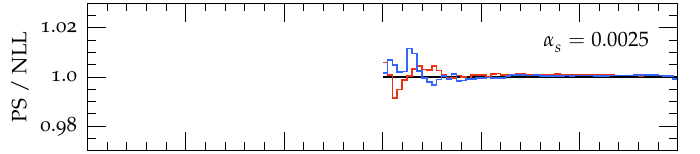}\\[-.8mm]
    \includegraphics[width=\textwidth]{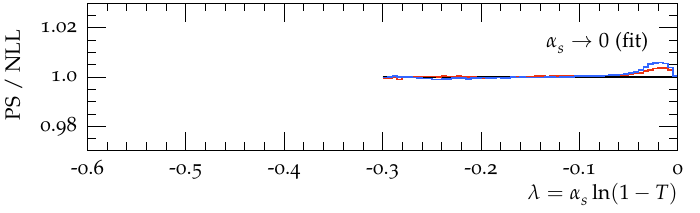}
  \end{minipage}\hfill
  \begin{minipage}{5.875cm}
    \includegraphics[width=\textwidth]{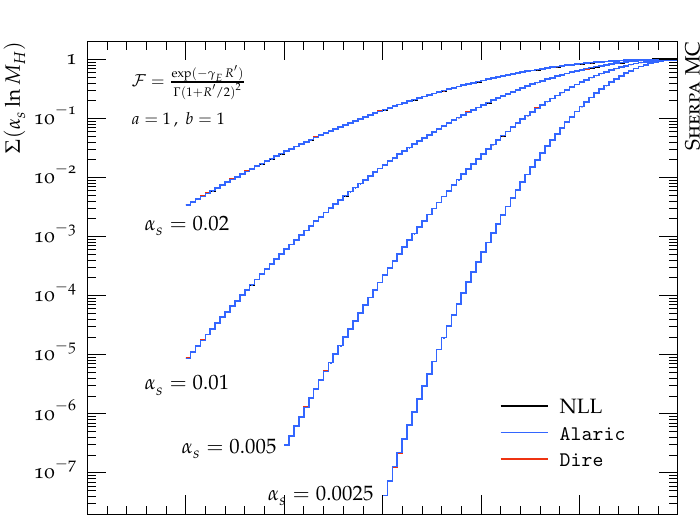}\\[-.8mm]
    \includegraphics[width=\textwidth]{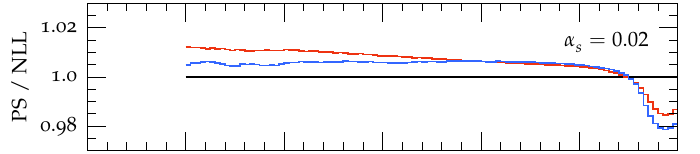}\\[-.8mm]
    \includegraphics[width=\textwidth]{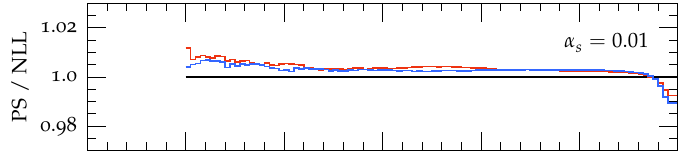}\\[-.8mm]
    \includegraphics[width=\textwidth]{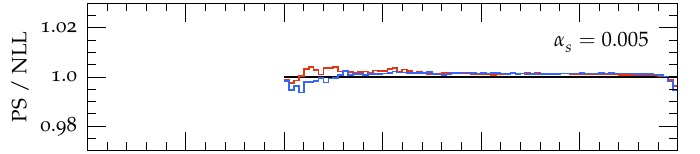}\\[-.8mm]
    \includegraphics[width=\textwidth]{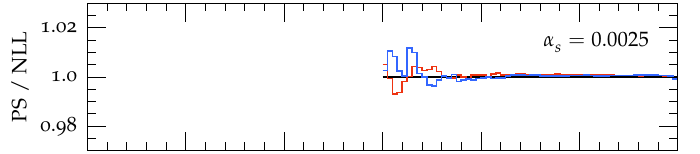}\\[-.8mm]
    \includegraphics[width=\textwidth]{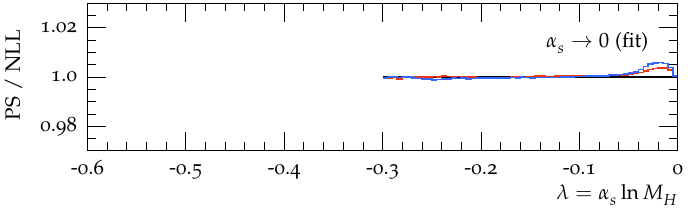}
  \end{minipage}\hfill
  \begin{minipage}{5.875cm}
    \includegraphics[width=\textwidth]{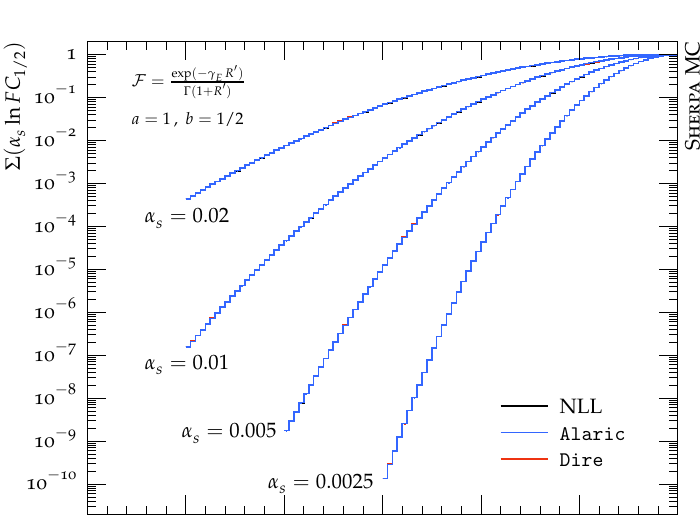}\\[-.8mm]
    \includegraphics[width=\textwidth]{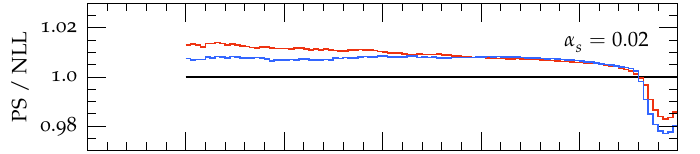}\\[-.8mm]
    \includegraphics[width=\textwidth]{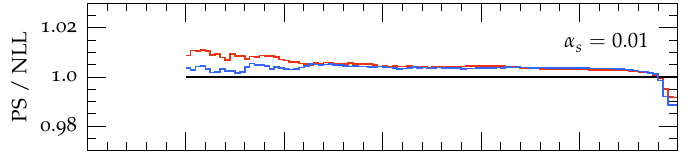}\\[-.8mm]
    \includegraphics[width=\textwidth]{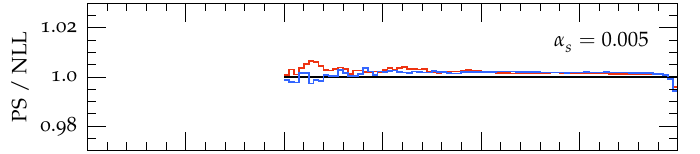}\\[-.8mm]
    \includegraphics[width=\textwidth]{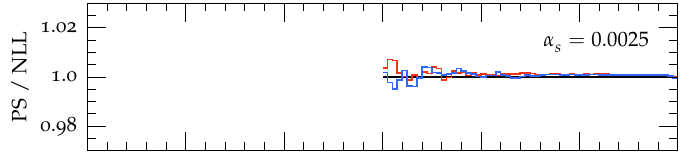}\\[-.8mm]
    \includegraphics[width=\textwidth]{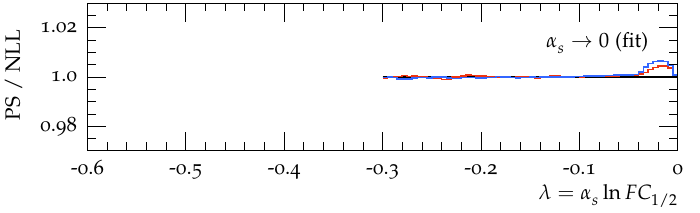}
  \end{minipage}
  \caption{NLL test for various event shape observables.
  See the main text for details.\label{fig:nll_test_shapes}}
\end{figure}

In this section we present numerical tests of our new algorithm\footnote{
    The PyPy code for these tests can be found at~\url{https://gitlab.com/shoeche/pyalaric}.}.
We follow the procedure outlined in~\cite{Dasgupta:2020fwr} and perform 
a scaling of the strong coupling, while keeping the variable 
$\lambda=\alpha_s\ln v$ fixed, where $v$ is an observable whose single-emission
contribution to a measurement can be parametrized in the form $v(k)=(k_t/Q)^a e^{-b|\eta_k|}$, 
see Eq.~\eqref{eq:V_approx}. In particular we analyze the event shape observables thrust, 
$T$~\cite{Farhi:1977sg}, jet broadening, $B_T$~\cite{Catani:1992jc}, heavy jet mass,
$M_H$, and the fractional energy correlators $FC_{1-\beta}$~\cite{Banfi:2004yd} for $\beta=0$ and $1/2$. 
We also analyze the leading Lund plane declustering scale in the Cambridge algorithm,
$y_{23}$, and the azimuthal angle between the two leading Lund plane declusterings, $\Delta\psi_{12}$~\cite{Dasgupta:2020fwr}.

Since the running of the strong coupling will not affect the kinematics reconstruction,
we keep $\alpha_s$ constant in this numerical test. In addition, we do not use 
the CMW scheme, and we work in the strict leading color approximation, $2C_F=C_A=3$. 
We find that this is sufficient to reproduce the dominant features of the \Dire 
dipole shower algorithm that were observed to break NLL precision in~\cite{Dasgupta:2018nvj,Dasgupta:2020fwr}. 
Figure~\ref{fig:nll_test_dpsi} shows the azimuthal angle separation $\Delta\psi_{12}$. 
The predictions from \Dire exhibit the same features as already shown in~\cite{Dasgupta:2020fwr}, 
and it can be seen that the deviation from a flat $\Delta\psi_{12}$ distribution 
does not vanish as $\alpha_s\to 0$. In contrast, for the \Alaric algorithm we observe 
increasingly smaller deviations from a flat $\Delta\psi_{12}$ dependence, 
in agreement with NLL resummation.

Figure~\ref{fig:nll_test_shapes} displays the event shape observables 
and the leading Lund declustering scale for varying $\alpha_s$. 
In order to test for a variety of possible effects of NLL violation, 
we have chosen observables with different NLL contributions. 
In addition, we test observables with $b=0$ ($\sqrt{y_{23}}$, $B_T$ and $FC_1$),
observables with $b=1/2$ ($FC_{1/2}$) and observables with $b=1$ ($1-T$, $M_H$). 
In each case we find that the deviation of the \Alaric prediction from the 
NLL target result (modified to account for constant $\alpha_s$, no CMW rescaling and leading color) 
decreases in size proportional to the scaling in $\alpha_s$, as $\alpha_s\to 0$.
At the same time, we observe large deviations of the \Dire predictions from the target NLL result.
It is notable that the predictions from \Alaric are flat with respect 
to the NLL result starting at fairly small values of $-\lambda$ for most observables.
For each prediction we have performed a fit to a linear function of $\alpha_s$
in order to extract the limit for $\alpha_s\to 0$. There are two noteworthy artefacts
of this extrapolation: Firstly, there are bumps in the extrapolated result at large
values of $\lambda$, which would not be present in the true ratio at any $\alpha_s<0.0025$.
Second, the extrapolated result is smoother than the individual inputs, since the
predictions at smaller $\alpha_s$ are less constraining due to their larger uncertainties.
This concludes our tests of the kinematics mapping.

\section{Comparison to experimental data}
\label{sec:results}
\begin{figure}[p]
  \centering
  \includegraphics[width=5.5cm]{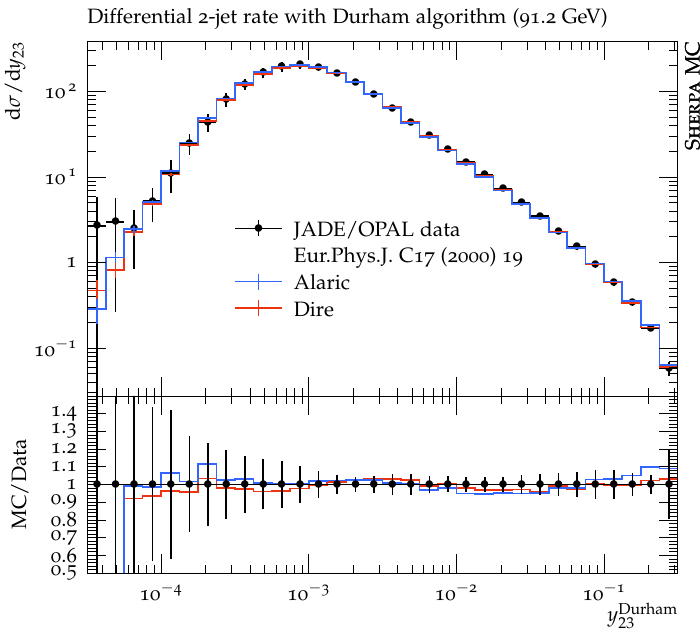}\hskip 5mm
  \includegraphics[width=5.5cm]{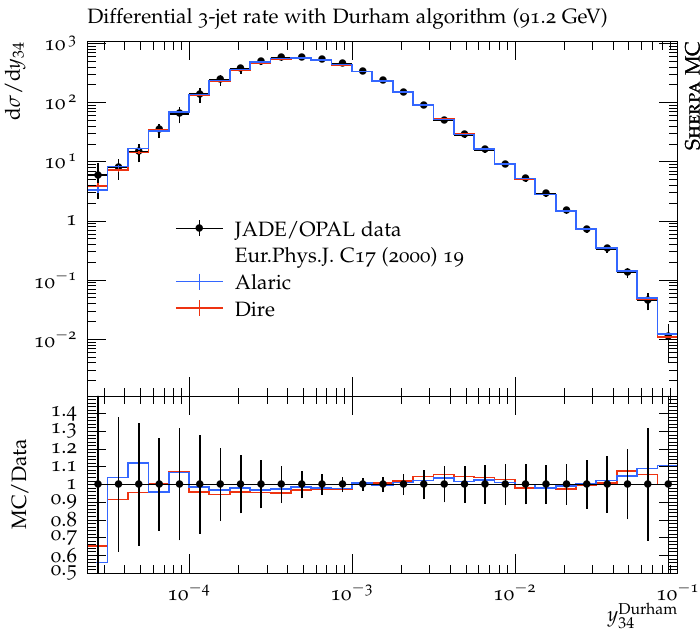}\\
  \includegraphics[width=5.5cm]{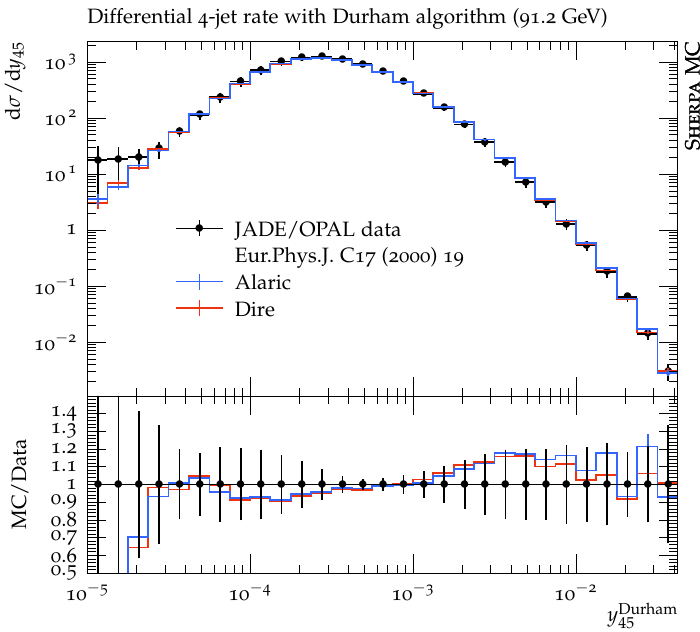}\hskip 5mm
  \includegraphics[width=5.5cm]{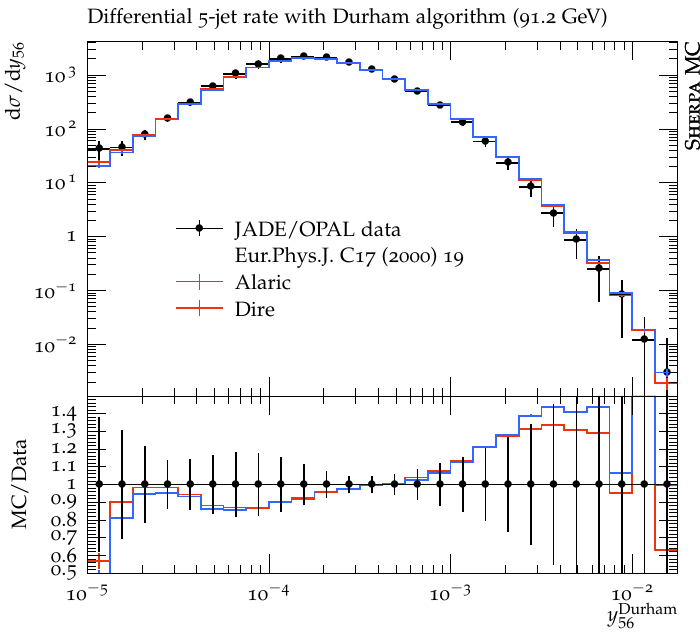}
  \caption{\Alaric and \Dire predictions in comparison to LEP data from~\cite{Pfeifenschneider:1999rz}.
    \label{fig:lep_jetrates}}
\end{figure}
\begin{figure}[p]
  \centering
  \includegraphics[width=5.5cm]{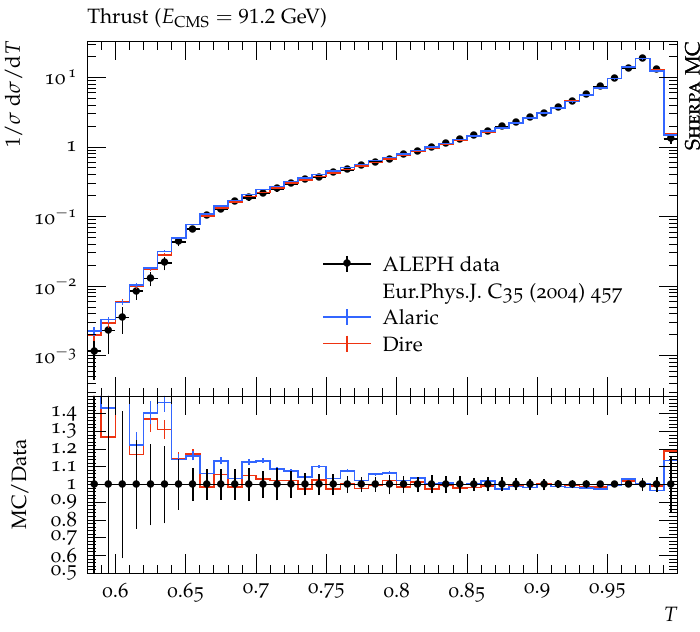}\hskip 5mm
  \includegraphics[width=5.5cm]{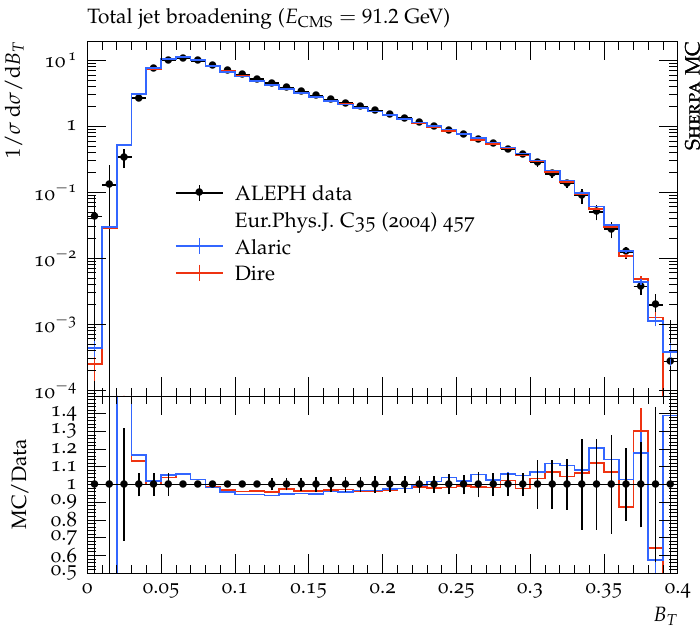}\\
  \includegraphics[width=5.5cm]{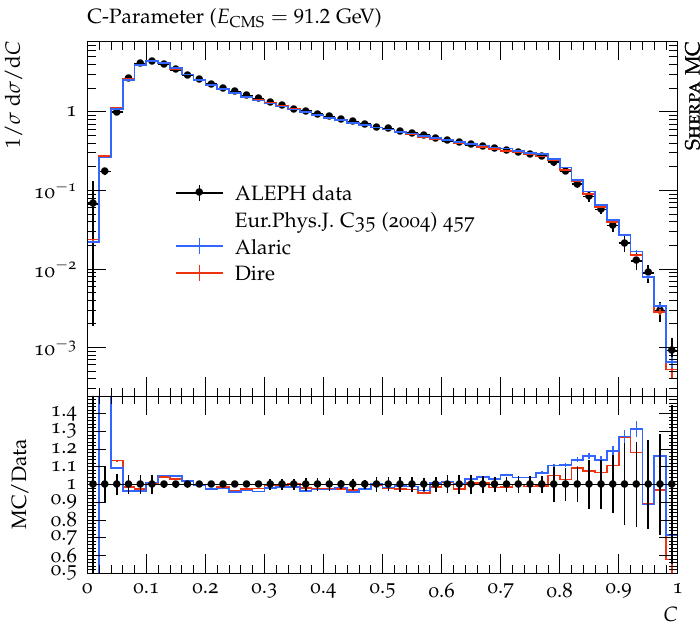}\hskip 5mm
  \includegraphics[width=5.5cm]{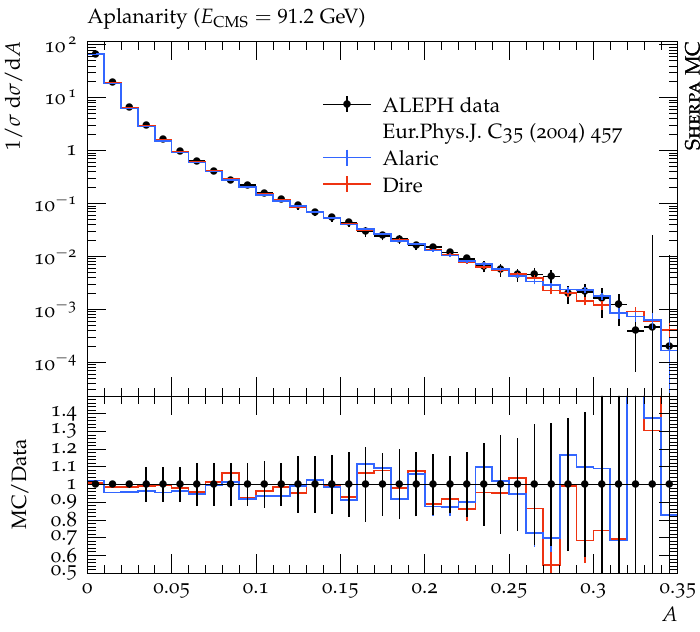}
  \caption{\Alaric and \Dire predictions in comparison to LEP data from~\cite{Heister:2003aj}.
    \label{fig:lep_shapes}}
\end{figure}
In this section we present first numerical results obtained with the \Alaric 
final-state parton shower, as implemented in the event generation framework
\Sherpa~\cite{Gleisberg:2003xi,Gleisberg:2008ta,Sherpa:2019gpd}. 
We do not perform NLO matching or multi-jet merging, and we set
$C_F=(N_c^2-1)/(2N_c)=4/3$ and $C_A=3$. All quarks are considered massless,
and we implement flavor thresholds at $m_c=1.42$~GeV and $m_b=4.92$~GeV.
The running coupling is evaluated at two loop accuracy, and we set $\alpha_s(m_z)=0.118$.
Following standard practice to improve the logarithmic accuracy of the parton shower,
we employ the CMW scheme~\cite{Catani:1990rr}, {\it i.e.}\ the soft eikonal contribution
to the flavor conserving splitting functions is rescaled by $1+\alpha_s(t)/(2\pi) K$, 
where $K=(67/18-\pi^2/6)\,C_A-10/9\,T_R\,n_f$.
Our results include the simulation of hadronization using the Lund string 
fragmentation implemented in Pythia~6.4~\cite{Sjostrand:2006za}~\footnote{
    Hadronization using the \Sherpa cluster fragmentation~\cite{Chahal:2022rid} 
    will require an implementation of massive splitting functions in \Alaric,
    in order to simulate the QCD evolution of partonic hadron decays. 
    We postpone this to a future publication.}.
We use the default hadronization parameters, apart from the following values:
\texttt{PARJ(21)=0.3}, \texttt{PARJ(41)=0.4}, \texttt{PARJ(42)=0.36} for \Alaric, and
\texttt{PARJ(21)=0.3}, \texttt{PARJ(41)=0.4}, \texttt{PARJ(42)=0.45} for \Dire.
All analyses are performed with Rivet~\cite{Buckley:2010ar}.

Figure~\ref{fig:lep_jetrates} shows predictions from the \Alaric parton shower for differential
jet rates in the Durham scheme compared to experimental results from the JADE and OPAL
collaborations~\cite{Pfeifenschneider:1999rz}. The perturbative region is to the right
of the plot, and $y\sim 2.8\cdot10^{-3}$ corresponds to the $b$-quark mass. The simulation
of nonperturbative effects dominates the predictions below $\sim10^{-4}$.
We observe fairly good agreement with the experimental data, however, we note that
at moderate values of $y$ the prediction will be altered when a proper evolution 
of massive quark final states is included.

Figure~\ref{fig:lep_shapes} shows a comparison for event shapes measured by the
ALEPH collaboration~\cite{Heister:2003aj}. The perturbative region is to the right
of the plot, except for the thrust distribution, where it is to the left.
We notice some deviation in the predictions for the total jet broadening and
for the aplanarity. They are mostly within the $2\sigma$ uncertainty
of the experimental measurements. It can be expected that the simulations will
improve upon including matrix-element corrections or when merging the \Alaric
parton shower with higher-multiplicity calculations. Deviations in the hadronization
region may be associated with the treatment of $b$-quarks and $c$-quarks as massless
partons.

\section{Conclusions}
\label{sec:summary}
We have presented a new parton-shower algorithm, which is closely modeled on the 
fixed-order subtraction formalism for identified particles by Catani and Seymour.
This technique allows, for the first time in a dipole-like parton shower,
to disentangle color and kinematics, at the price of introducing an azimuthal
angle dependence in the splitting functions. Partial fractioning the angular radiator
function and matching to the collinear limit maintains strict positivity
of the evolution kernels, thus allowing a straightforward implementation without the need
for explicit angular ordering. Through a suitable assignment of recoil, which is
absorbed by the entire QCD multipole, we are able to show that the local 
kinematics mapping satisfies the criteria for NLL precision.

Several extensions of this algorithm are required: 
Firstly, it should be modified to include spin correlations~\cite{
  Collins:1987cp,Knowles:1987cu,Knowles:1988vs,Knowles:1988hu}
and dominant sub-leading color effects~\cite{Gustafson:1992uh,Dulat:2018vuy,Hamilton:2020rcu}.
A number of formally less relevant, but practically important considerations
need to be addressed as well. They include the evolution for massive partons,
in order to properly describe bottom and charm jet production and $b$- and
$c$-quark fragmentation functions. Another open question is the extension to
processes with non-trivial color dependence at Born level, such as top-quark
pair production and inclusive jet or di-jet production at hadron colliders.
A related, though substantially simpler problem is the treatment of processes
with multiple, disconnected QCD multipoles, such as vector boson fusion in the
structure function approximation, or the production of a Higgs boson in association
with a hadronically decaying vector boson in the narrow width approximation.
The latter cases can be handled by applying the algorithms introduced here
to each QCD multipole individually, while keeping track of spin correlations
among the different multipoles.
Finally, we plan to extend our new algorithms to higher-orders, based on the
techniques developed in~\cite{Hoche:2017iem,Dulat:2018vuy,Gellersen:2021eci}.

\section*{Acknowledgments}
We thank Andrea Banfi for pointing out that the relative difference between the
parton-shower and the NLL result in Fig.~\ref{fig:nll_test_shapes}
should decrease proportional to $\alpha_s$ as $\alpha_s\to 0$.
The work of F.~Herren and S.~H{\"o}che was supported by the
Fermi National Accelerator Laboratory (Fermilab),
a U.S. Department of Energy, Office of Science, HEP User Facility.
Fermilab is managed by Fermi Research Alliance, LLC (FRA),
acting under Contract No. DE--AC02--07CH11359.
F.~Herren acknowledges support by the Alexander von Humboldt foundation.
This work has received funding from the European Union’s Horizon 2020
research and innovation programme as part of the Marie Sk{\l}odowska-Curie
Innovative Training Network MCnetITN3 (grant agreement no.~722104).
F.~Krauss acknowledges funding as Royal Society Wolfson Research fellow.
F.~Krauss and M.~Sch{\"o}nherr are supported by the STFC under grant agreement ST/P001246/1.
They would like to thank the Munich Institute for
Astro-, Particle and BioPhysics (MIAPbP), funded by the Deutsche Forschungsgemeinschaft
(DFG, German Research Foundation) under Germany's Excellence Strategy
– EXC-2094 – 390783311, for hospitality while this publication was finalized.
The work of M.~Sch{\"o}nherr was supported by the Royal Society through a
University Research Fellowship (URF\textbackslash{}R1\textbackslash{}180549)
and an Enhancement Award (RGF\textbackslash{}EA\textbackslash{}181033,
CEC19\textbackslash{}100349, and RF\textbackslash{}ERE\textbackslash{}210397).

\appendix
\section{Implementation details}
\label{sec:mc_details}
This appendix summarizes details of the kinematics mapping and the relations between the kinematic
invariants and the evolution and splitting variables for both final-state and initial-state evolution. We limit the discussion to situations where the recoiler momentum
is composed either of final-state momenta only, or of initial-state momenta only.
\subsection{Final-state evolution}
\label{sec:fs_ps}
We will first discuss the case of a final-state emitter with a recoil momentum $\tilde{K}$ that is composed only of final-state momenta.
The momentum mapping is sketched in Fig.~\ref{fig:kinematics}~(FF). The emitting particle is labeled
$i$, the emission is labeled $j$, and the color spectator is labeled $k$.
The momenta $p_i$, $p_j$ and $K$ are given by Eq.~\eqref{eq:pi_K_using_kt2}
\begin{equation}
\label{eq:ff_emit_spec}
  \begin{split}
    p_i&=\,z\,\tilde{p}_i\;,\\
    p_j&=\,(1-z)\,\tilde{p}_i+v\big(\tilde{K}-(1-z+2\kappa)\,\tilde{p}_i\big)+k_\perp\;,\\
    K&=\,\tilde{K}-v\big(\tilde{K}-(1-z+2\kappa)\,\tilde{p}_i\big)-k_\perp\;,
  \end{split}
\end{equation}
where
\begin{equation}
  {\rm k}_\perp^2=v(1-v)(1-z)\,2\tilde{p}_i\tilde{K}-v^2\tilde{K}^2\;,
  \qquad\text{and}\qquad
  v=\frac{\tau}{1-z}\;.
\end{equation}
The particles included in the momentum $\tilde{K}$ are subjected to a Lorentz transformation,
which accounts for the decay of the new recoil momentum, $K$, in a different frame. If the recoil momentum
is given by a single, light-like vector, no such transformation is necessary.
\begin{equation}\label{eq:lt_ff}
    p_l^\mu\to\Lambda^\mu_{\;\nu}(\tilde{K},K)\,p_l^\nu\;.
\end{equation}
The momentum mapping for final-state emitters with with a recoil momentum
$\tilde{K}$ composed only of initial-state momenta is sketched 
in Fig.~\ref{fig:kinematics}~(FI). We define the momentum $\tilde{K}$ as incoming, 
i.e.\ $\tilde{K}_0>0$. This implies the replacement $v\to -v$, $\tilde{K}\to -\tilde{K}$ 
and $n\to -n$, leading to
\begin{equation}
\label{eq:fi_emit_spec}
  \begin{split}
    p_i&=\,z\,\tilde{p}_i\;,\\
    p_j&=\,(1-z)\,\tilde{p}_i+v\big(\tilde{K}+(1-z-2\kappa)\,\tilde{p}_i\big)+k_\perp\;,\\
    K&=\,\tilde{K}+v\big(\tilde{K}+(1-z-2\kappa)\,\tilde{p}_i\big)+k_\perp\;,
  \end{split}
\end{equation}
where
\begin{equation}
  {\rm k}_\perp^2=v(1+v)(1-z)\,2\tilde{p}_i\tilde{K}-v^2\tilde{K}^2\;,
  \qquad\text{and}\qquad
  v=\frac{\tau}{1-z}\;.
\end{equation}
As in the case of a final-state spectator, the momenta defining $\tilde{K}$ are subjected
to the Lorentz transformation given by Eq.~\eqref{eq:lt_ff}.
An additional transformation has to be applied, such as to align the momenta 
of both redefined beam particles, $p_a$ and $p_b$, with the beam axis.
If there are no strongly interacting particles in the initial state, 
this can be achieved by the simple mapping
\begin{equation}\label{eq:lt_fi}
  p_l^\mu\to\Lambda^\mu_{\;\nu}(p_a+p_b,\tilde{p}_a+\tilde{p}_b)\,p_l^\nu\;.
\end{equation}
Note that this additional Lorentz transformation is applied to all particles in the event. 

\subsection{Initial-state evolution}
\label{sec:is_ps}
\begin{figure}[t]
\includegraphics[width=\textwidth]{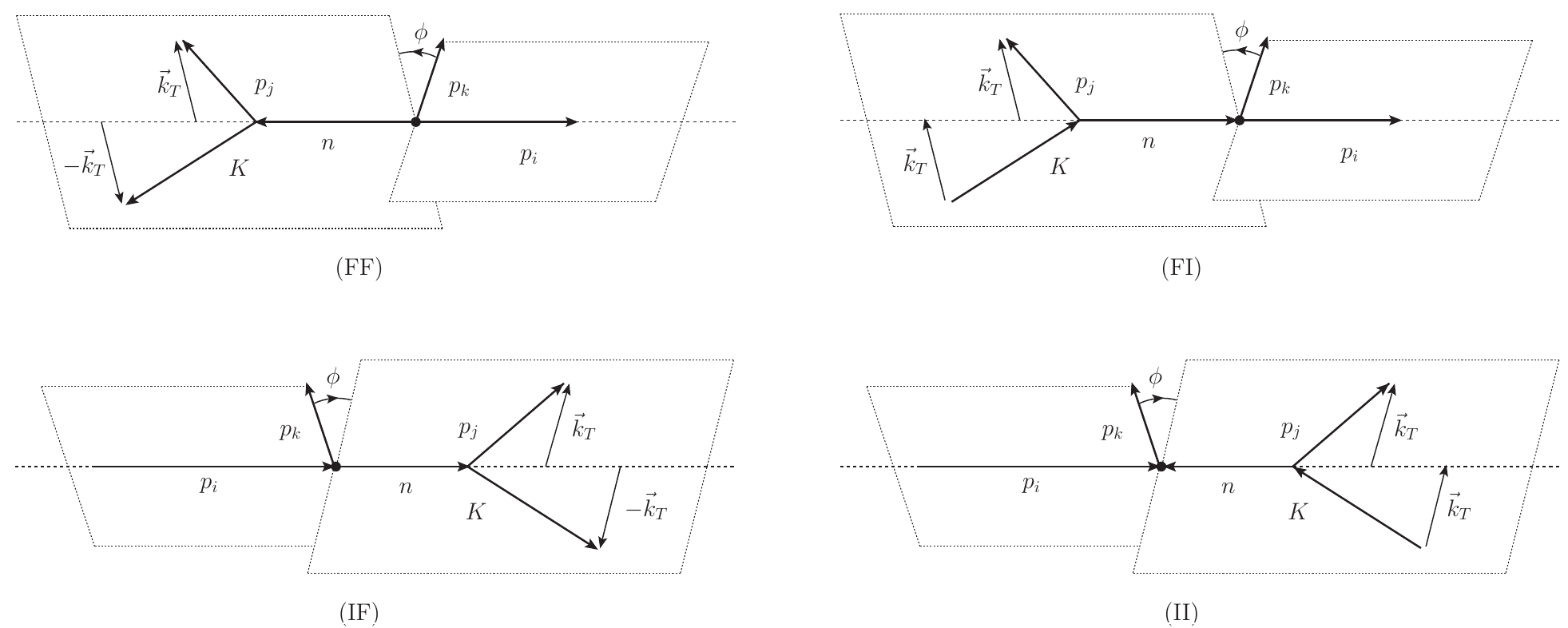}
\caption{Sketch of the momentum mapping
for final-state emitters with final-state spectator~(FF),
final-state emitters with initial-state spectator~(FI),
initial-state emitters with final-state spectator~(IF)
and initial-state emitters with initial-state spectator~(II).
\label{fig:kinematics}}
\end{figure}
For initial-state emissions, we redefine $z\to 1/z$. The momentum mapping for
recoil momenta $\tilde{K}$ composed only of initial-state momenta
is sketched in Fig.~\ref{fig:kinematics}~(II). The emitting particle is labeled
$i$, the emission is labeled $j$, and the color spectator is labeled $k$. We define both
$\tilde{p}_i$ and $\tilde{K}$ as incoming, i.e.\ $\tilde{p}_{i,0}>0$ and $\tilde{K}_0>0$.
The momenta $p_i$, $p_j$ and $K$ are then given by Eq.~\eqref{eq:pi_K_using_kt2}
\begin{equation}
\label{eq:ii_emit_spec}
  \begin{split}
    p_i&=\,\frac{1}{z}\,\tilde{p}_i\;,\\
    p_j&=\,\frac{1-z}{z}\,\tilde{p}_i+v\left(\tilde{K}+\left(\frac{1-z}{z}-2\kappa\right)\tilde{p}_i\right)+k_\perp\;,\\
    K&=\,\tilde{K}+v\left(\tilde{K}+\left(\frac{1-z}{z}-2\kappa\right)\tilde{p}_i\right)+k_\perp\;,
  \end{split}
\end{equation}
where
\begin{equation}
  {\rm k}_\perp^2=v(1+v)\frac{1-z}{z}\,2\tilde{p}_i\tilde{K}-v^2\tilde{K}^2\;,
  \qquad\text{and}\qquad
  v=\frac{\tau}{1-z}\;.
\end{equation}
As in the case of final-state evolution, the momenta defining $\tilde{K}$ are subjected
to the Lorentz transformation given by Eq.~\eqref{eq:lt_ff}.
The complete event is then subjected to a Lorentz transformation, determined such as to 
align the momenta of the initial-state particles, $p_a$ and $p_b$, with the beam axis,
while shifting the event rapidity to $y=\tilde{y}-{\rm sgn}(p_{i,z})\ln\sqrt{z}$
and preserving the azimuthal orientation of the event.

The momentum mapping for initial-state emitters with a recoil momentum 
$\tilde{K}$ composed only of final-state momenta is sketched 
in Fig.~\ref{fig:kinematics}~(IF). The momenta $p_i$, $p_j$ and $K$ are given by
\begin{equation}
\label{eq:if_emit_spec}
  \begin{split}
    p_i&=\,\frac{1}{z}\,\tilde{p}_i\;,\\
    p_j&=\,\frac{1-z}{z}\,\tilde{p}_i+v\left(\tilde{K}-\left(\frac{1-z}{z}+2\kappa\right)\tilde{p}_i\right)+k_\perp\;,\\
    K&=\,\tilde{K}-v\left(\tilde{K}-\left(\frac{1-z}{z}+2\kappa\right)\tilde{p}_i\right)-k_\perp\;,
  \end{split}
\end{equation}
where
\begin{equation}
  {\rm k}_\perp^2=v(1-v)\frac{1-z}{z}\,2\tilde{p}_i\tilde{K}-v^2\tilde{K}^2\;,
  \qquad\text{and}\qquad
  v=\frac{\tau}{1-z}\;.
\end{equation}
The particles included in the momentum $\tilde{K}$ are subjected to a Lorentz transformation,
which accounts for the decay of the new recoil momentum, $K$, in a different frame.
As in the case of final-state evolution, this can be achieved by applying 
the Lorentz transformation in Eq.~\eqref{eq:lt_ff}. If the recoil momentum
is given by a single, light-like vector, no transformation is needed.

\section{Phase-space factorization}
\label{sec:ps_factorization}
This appendix summarizes details on the phase-space factorization 
for both final-state and initial-state evolution. The case of 
final-state emitter and initial-state recoiler was discussed
in Sec.~\ref{sec:kinematics}, and we do not repeat it here.

\subsection{Final-state emitter and final-state recoiler}
This case covers electroweak decays of a color-charged resonance,
such as the top quark. We start from
\begin{equation}
    {\rm d}\Phi_{3}(p_i,p_j,K;Q)
    ={\rm d}\Phi_{2}(p_i,n;Q)\,
    \frac{{\rm d}n^2}{2\pi}\,
    {\rm d}\Phi_{2}(p_j,K;n)
\end{equation}
The generic frame-independent form of the two-particle phase space
is given in Eq.~\eqref{eq:two_body_ps}. We perform all transformations
in the rest frame of $n$, where we have the simple relations
\begin{equation}\label{eq:energies_n_frame_fs}
    E_i=z\,\frac{\tilde{p}_i\tilde{K}}{\sqrt{n^2}}\;,\qquad
    E_j=(1-z)\,\frac{\tilde{p}_i\tilde{K}}{\sqrt{n^2}}\;,\qquad
    E_K=(1-z+2\kappa)\,\frac{\tilde{p}_i\tilde{K}}{\sqrt{n^2}}\;,
    \qquad\text{and}\qquad
    n^2=2\tilde{p}_i\tilde{K}\,(1-z+\kappa)\;.
\end{equation}
Using the following identity for the polar angle $\theta_j$ of the emission,
\begin{equation}\label{eq:ct_nframe_fs}
    1-\cos\theta_j^{\,i}=-2v\,\frac{1-z+\kappa}{1-z}\;,
\end{equation}
we find the first two-particle decay phase space to be
\begin{equation}\label{eq:ndecay_nframe_fs}
    {\rm d}\Phi_{2}(p_j,K;n)=
    \frac{1}{16\pi^2}\,{\rm d}v\,\frac{{\rm d}\phi_j^{(n)}}{2\pi}\;.
\end{equation}
Analogous to Eq.~\eqref{eq:n2_trafo_fi}, we can write 
${\rm d}n^2=2\tilde{p}_i\tilde{K}\,{\rm d}z$.
Finally, we rewrite the second two-particle decay phase space
\begin{equation}\label{eq:ps_fact_ff_ndecay}
    {\rm d}\Phi_{2}(p_i,n;Q)=
    \frac{1}{16\pi^2}\frac{z}{2(1-z+\kappa)}\,
    {\rm d}\cos\theta_i^{(n)}\,\frac{{\rm d}\phi_i^{(n)}}{2\pi}\;.
\end{equation}
In order to obtain a factorization formula, this must be mapped
to the Born phase space, which is given by
${\rm d}\Phi_{2}(\tilde{p}_i,\tilde{K};Q)$.
The angular integrals in Eq.~\eqref{eq:ps_fact_ff_ndecay}
are identical when working in the rest frame of the momentum $n$,
which leads to the relation
${\rm d}\Phi_{2}(p_i,n;Q)=
  z\;{\rm d}\Phi_{2}(\tilde{p}_i,\tilde{q}_i;\tilde{K})$.
Combining all of the above, we find the single-emission 
phase space element
\begin{equation}
    {\rm d}\Phi_{+1}^{\rm(FF)}(\tilde{K};\tilde{p}_1,\ldots,
    \tilde{p}_{j-1},\tilde{p}_{j+1},\ldots,\tilde{p}_n;p_j)
    =\frac{2z\tilde{p}_i\tilde{K}}{16\pi^2}\,
    {\rm d}v\,{\rm d}z\,\frac{{\rm d}\phi}{2\pi}\;.
\end{equation}

\subsection{Initial-state emitter and final-state recoiler}
This case covers hadroproduction of a colorless final state, in particular 
Drell-Yan lepton pair production. We start from the factorization formula
\begin{equation}
    {\rm d}\Phi_{2}(p_j,K;Q+p_i)
    ={\rm d}\Phi_{2}(p_j,K;n)\,
    \frac{{\rm d}n^2}{2\pi}\,
    {\rm d}\Phi_{1}(n;Q+p_i)
\end{equation}
The generic frame-independent form of the two-particle phase space
is given in Eq.~\eqref{eq:two_body_ps}. Again, we perform all transformations
in the rest frame of $n$, leading to the relations 
in Eq.~\eqref{eq:energies_n_frame_fs}.
Using Eq.~\eqref{eq:ct_nframe_fs} for the polar angle $\theta_j$ of the emission,
we find the two-particle decay phase space in Eq.~\eqref{eq:ndecay_nframe_fs}.
Analogous to Eq.~\eqref{eq:n2_trafo_fi}, we can write 
${\rm d}n^2=2\tilde{p}_i\tilde{K}\,{\rm d}z$.
The one-particle production phase space is given by
\begin{equation}
  {\rm d}\Phi_{1}(n;Q+p_i)=
  2\pi\,\delta(n^2-(Q+p_i)^2)\;.
\end{equation}
In order to obtain a factorization formula, this is mapped
to the Born phase space, leading to ${\rm d}\Phi_{1}(n;Q+p_i)=
  1/z\;{\rm d}\Phi_{2}(\tilde{K};Q+\tilde{p}_i)$.
Combining all of the above, we obtain the single-emission 
phase space element
\begin{equation}
    {\rm d}\Phi_{+1}^{\rm(IF)}(\tilde{K};\tilde{p}_1,\ldots,
    \tilde{p}_{j-1},\tilde{p}_{j+1},\ldots,\tilde{p}_n;p_j)
    =\frac{2\tilde{p}_i\tilde{K}}{16\pi^2}\,
    {\rm d}v\,\frac{{\rm d}x}{x}\,\frac{{\rm d}\phi}{2\pi}\;.
\end{equation}

\subsection{Initial-state emitter and initial-state recoiler}
This case covers deep inelastic scattering. We start from the two-particle phase space
\begin{equation}
    {\rm d}\Phi_{2}(p_j,Q;K+p_i)\;.
\end{equation}
Its generic, frame-independent form is given in Eq.~\eqref{eq:two_body_ps}.
We perform all transformations in the rest frame of $n$, where we have the
relations in Eq.~\eqref{eq:energies_n_frame_is}. Next we insert the identity
\begin{equation}
  1=\int{\rm d}x\;\delta\left(x-\frac{\tilde{p}_i\tilde{K}}{p_in}\right)
  =\int{\rm d}x\;\frac{2\tilde{p}_i\tilde{K}}{x}\,
  \delta\left(2\tilde{p}_i\tilde{K}-(p_i+p_j)n\right)
  =\int{\rm d}x\;\frac{2\tilde{p}_i\tilde{K}}{x}\,
  \delta\left((\tilde{p}_i+\tilde{K})^2-Q^2\right)\;.
\end{equation}
In order to obtain a factorization formula, this is mapped to the Born phase space as
${\rm d}\Phi_{1}(\tilde{Q};\tilde{p}_i+\tilde{K})=2\pi\,\delta(\tilde{Q}^2-(\tilde{p}_i+\tilde{K})^2)$.
Combining all of the above, we obtain the single-emission 
phase space element
\begin{equation}
    {\rm d}\Phi_{+1}^{\rm(IF)}(\tilde{K};\tilde{p}_1,\ldots,
    \tilde{p}_{j-1},\tilde{p}_{j+1},\ldots,\tilde{p}_n;p_j)
    =\frac{2\tilde{p}_i\tilde{K}}{16\pi^2}\,
    {\rm d}v\,\frac{{\rm d}x}{x}\,\frac{{\rm d}\phi}{2\pi}\;.
\end{equation}

\section{Monte-Carlo counterterms for MC@NLO matching}
\label{sec:nlo matching}
To match a parton shower to NLO calculations in dimensional regularization based on the 
MC@NLO algorithm \cite{Frixione:2002ik}, the integral of the splitting functions must be known
in $D=4-2\eps$ dimensions. Since our new parton shower is modeled on the Catani-Seymour identified
particle subtraction, we can utilize the techniques developed in~\cite{Hoche:2018ouj,Liebschner:2020aa}.
By means of a suitable identification of the radiating color multipole, this allows us to treat 
the problem of standard final-state evolution, resonant particle production, color singlet production
at hadron colliders, deep-inelastic scattering, etc. on the same footing.

Here we will limit ourselves to listing the main changes with respect to~\cite{Hoche:2018ouj,Liebschner:2020aa}
that are needed in order to implement the algorithm. For details on the respective phase-space integrals,
and on the basis of the subtraction technique, we refer the reader to~\cite{Catani:1996vz}. Details on
the implementation of MC@NLO can be found in~\cite{Frixione:2002ik,Hoeche:2011fd}.

\subsection{Final-state emitter}
By combining the integrated splitting function with the collinear mass factorization counterterms,
we can derive a combined integrated subtraction term for identified parton production 
with a partonic fragmentation function~\cite{Hoche:2018ouj,Liebschner:2020aa}
\begin{equation}\label{eq:sigmaI}
    \int_{m+1}\mathrm{d}\sigma^S + \int_m\mathrm{d}\sigma^C 
    = \frac{1}{2}\sum_{i=g,q,\bar{q}}
    \sum_{\tilde{\imath}=1}^m\int_0^1 \frac{\mathrm{d} z}{ z^{2-2\epsilon} }
    \int_m \mathrm{d}\sigma^B(p_1,\dots,\frac{p_i}{z},\dots,p_m)\otimes
    \hat{\mathbf{I}}_{\tilde{\imath}i}^{\rm(FS)}\;,
\end{equation}
where the $\otimes$ stands for spin and color correlations.
In the $\overline{\mathrm{MS}}$ scheme, the insertion operator is given by
\begin{equation}
  \hat{\mathbf{I}}_{\tilde{\imath}i}^{\rm(FS)}
  = - \frac{\alpha_s}{2\pi}\frac{1}{\Gamma(1-\epsilon)}\Bigg\{ \sum_{k=1,k\neq\tilde{\imath}}^m
  \frac{\mathbf{T}_{\tilde{\imath}}\mathbf{T}_{k}}{\mathbf{T}_{\tilde{\imath}}^2}
  \left(\frac{4\pi\mu^2(p_kn)}{2(p_i p_k)(p_in)} \right)^{\epsilon} 
    \hat{\mathcal{V}}_{\tilde{\imath},i}(z; \epsilon ; p_i, p_k, n) - \delta_{\tilde{\imath}i}
    \frac{1}{\epsilon}\left(\frac{4\pi\mu^2}{\mu_F^2}\right)^\epsilon P_{\tilde{\imath} i}(z)\Bigg\}~.
\end{equation}
The explicit pole in $\epsilon$, originating from the renormalization of the
perturbative fragmentation function, cancels against the corresponding pole in
$\bar{\mathcal{V}}_{\tilde{\imath},i}$. The remainder can be split into three contributions:
\begin{equation}
    \hat{\mathbf{I}}_{\tilde{\imath}i}^{\rm(FS)} = 
    \delta(1-z) \mathbf{I}_{\tilde{\imath}i} + \mathbf{P}_{\tilde{\imath}i} + \mathbf{H}_{\tilde{\imath}i}~.
\end{equation}
The singularities in the virtual corrections are canceled by the insertion operator present for standard
final-state dipoles with final-state spectator~\cite{Catani:1996vz}
\begin{equation}\label{eq:i_operator}
  \mathbf{I}_{\tilde{\imath}i}(p_1,\dots,p_i,\dots,p_m;\epsilon)
  = - \frac{\alpha_s}{2\pi}\frac{1}{\Gamma(1-\epsilon)}\sum_{k=1,k\neq\tilde{\imath}}^m
  \frac{\mathbf{T}_{\tilde{\imath}}\mathbf{T}_{k}}{\mathbf{T}_{\tilde{\imath}}^2}
  \left(\frac{4\pi\mu^2(p_kn)}{2(p_i p_k)(p_in)}\right)^{\epsilon}\mathcal{V}_{\tilde{\imath}i}(\epsilon)\;.
\end{equation}
Employing color conservation and expanding through $\mathcal{O}(\epsilon)$, the remaining two operators
read \cite{Hoche:2018ouj,Liebschner:2020aa}
\begin{equation}
  \mathbf{P}_{\tilde{\imath}i}(p_1,\dots,\frac{p_i}{z},\dots,p_m;z;\mu_F)
  = \frac{\alpha_s}{2\pi}\sum_{k=1,k\neq\tilde{\imath}}^m
  \frac{\mathbf{T}_{\tilde{\imath}}\mathbf{T}_{k}}{\mathbf{T}_{\tilde{\imath}}^2}
  \ln\frac{z^2\mu_F^2(p_kn)}{2(p_i p_k)(p_in)}\delta_{\tilde{\imath} i}P_{\tilde{\imath} i}(z)
\end{equation}
and
\begin{equation}\label{eq:h_operator_cs}
  \mathbf{H}_{\tilde{\imath}i}(p_1,\dots,p_i,\dots,p_m;n;z)
  = -\frac{\alpha_s}{2\pi}\sum_{k=1,k\neq\tilde{\imath}}^m
  \frac{\mathbf{T}_{\tilde{\imath}}\mathbf{T}_{k}}{\mathbf{T}_{\tilde{\imath}}^2}
  \left[\tilde{K}^{\tilde{\imath} i}(z) + \bar{K}^{\tilde{\imath} i}(z)
    + 2 P_{\tilde{\imath} i}(z)\ln z + \hat{\mathcal{L}}^{\tilde{\imath} i}(z; p_i, p_k, n) \right]~.
\end{equation}
Finally, an integration over $p_i$ has to be performed. Following Refs.~\cite{Hoche:2018ouj,Liebschner:2020aa}
we replace the integration over $p_i$ by an integration over $\tilde{p}_i$. This leads to a Jacobian of $z^{2-2\epsilon}$,
canceling the prefactor in Eq.~\eqref{eq:sigmaI}. Consequently, the differential Born cross-section 
$\mathrm{d}\sigma^B$ decouples from the $z$ integration and we obtain
\begin{equation}
    \sum_{\tilde{\imath}=1}^m\int\frac{\mathrm{d}^D p_i}{(2\pi)^{D-1}}\delta(p_i^2)\sigma^I(p_i)
    = \sum_{\tilde{\imath}=1}^m\int_m\mathrm{d}\sigma^B(p_1,\dots,p_m)\otimes
    \int_0^1\mathrm{d}z\;\frac{1}{2}\sum_{i=g,q,\bar{q}}\hat{\mathbf{I}}_{\tilde{\imath}i}\;.
\end{equation}
The integral over $\sum_i\mathbf{P}_{\tilde{\imath}i}$ vanishes, because $\sum_i P_{\tilde{i}i}$ is
a pure plus distribution. The other two $z$-integrals can be evaluated directly, because $\tilde{p}_i$
is $z$-independent. In particular, the integral of the $\mathbf{I}_{\tilde{\imath}}$ operator is trivial.
The integral over the $\mathbf{H}_{\tilde{\imath}i}$ operator is given by a modified form
of the result in~\cite{Hoche:2018ouj,Liebschner:2020aa}. We note that the arguments 
of the dilogarithms depend only on the angle and velocity of $l_{ik}$ as defined
in Eq.~\eqref{eq:frame_independent_radiator}, see~Appendix~B in~\cite{Catani:1996vz}.
Using the following substitution in Eq.~(B.9) in~\cite{Catani:1996vz}\footnote{
  Note that in this context $v$ is defined as the relative velocity of $l_{ik}$ and $n$,
  see Appendix~B in~\cite{Catani:1996vz}.}
\begin{equation}
  1-v\cos\chi\to\frac{n^2\,l_{ik}p_i}{(p_in)(l_{ik}n)}\;,
  \qquad
  1-v^2\to\frac{n^2l_{ik}^2}{(l_{ik}n)^2}\;,
\end{equation}
we therefore obtain
\begin{equation}\label{eq:h_operator}
\begin{split}
    &\int_0^1\mathrm{d}{z}\, \mathbf{H}_{\tilde{\imath}i}(p_1,\dots,p_i,\dots,p_m;n;z)\\
    &\qquad=- \frac{\alpha_s}{2\pi}\sum_{k=1,k\neq\tilde{\imath}}^m
    \frac{\mathbf{T}_{\tilde{\imath}}\mathbf{T}_{k}}{\mathbf{T}_{\tilde{\imath}}^2}
    \Bigg\{\;\mathcal{K}^{\tilde{\imath}i}
    +\delta_{\tilde{\imath}i}\,\mathrm{Li}_2\bigg(1-
    \frac{2\tilde{p}_i\tilde{p}_k\,\tilde{K}^2}{(\tilde{p}_i\tilde{K})(\tilde{p}_k\tilde{K})}\bigg)
    -\int_0^1\mathrm{d}z\,P^{\tilde{\imath}i}_{\rm reg}(z)
    \ln\frac{(p_ip_k)n^2}{2(p_i n)(p_kn)}\;\Bigg\}\;,
    \end{split}
\end{equation}
where the integral $\mathcal{K}^{\tilde{\imath}i}$ is defined as
\begin{equation}
  \mathcal{K}^{\tilde{\imath}i} = \int_0^1\mathrm{d}z~\Big(\bar{K}^{\tilde{\imath} i}(z)
  +\tilde{K}^{\tilde{\imath} i}(z)+2 P_{\tilde{\imath} i}(z)\ln z\Big)\;.
\end{equation}
In general, the last term of Eq.~\eqref{eq:h_operator} must be computed numerically,
as $n$ implicitly depends on $z$, see Eq.~\eqref{eq:def_n_pi}.

\subsection{Initial-state emitter}
As in the case of a final-state emitter, the case of an initial state emitter is treated 
in the same manner as in \cite{Hoche:2018ouj,Liebschner:2020aa}. The sum of the subtraction terms
and the collinear counterterms is given by
\begin{equation}
    \int_{m+1}\mathrm{d}\sigma^S + \int_m\mathrm{d}\sigma^C = 
    \sum_{\tilde{i}=g,q,\bar{q}}\int_0^1\mathrm{d}x
    \int_{m}\mathrm{d}\sigma^B(p_1,\dots,x p_i,\dots,p_m)\otimes
    \hat{\mathbf{I}}_{i\tilde{\imath}}^{\rm(IS)}\;,
\end{equation}
where the $\otimes$ again stands for spin and color correlations. In the $\overline{\rm MS}$ scheme,
the insertion operator is given by
\begin{equation}
    \hat{\mathbf{I}}_{i\tilde{\imath}}^{\rm(IS)} = 
    \delta(1-z) \mathbf{I}_{i\tilde{\imath}} + \mathbf{P}_{i\tilde{\imath}} + \mathbf{K}_{i\tilde{\imath}}\;.
\end{equation}
The operator $\mathbf{I}_{i\tilde{\imath}}$ is obtained by replacing $k\to\tilde{k}$ 
in Eq.~\eqref{eq:i_operator}. The $\mathbf{K}$ operator can be written as
\begin{align}
  \mathbf{K}_{i\tilde{\imath}}(p_1, \hdots, p_m; p_i, n, x) 
  = -\frac{\alpha_s}{2\pi} \sum_{\tilde{b}=1}^m 
  \frac{\mathbf{T}_{\tilde{k}}\mathbf{T}_{\tilde{\imath}}}{\mathbf{T}_{\tilde{\imath}}^2}
  \left[\; \bar{K}^{i,\tilde{\imath}}(x) + \tilde{K}^{i,\tilde{\imath}}(x)
    + \mathcal{L}^{i,\tilde{\imath}}(x;p_i,\tilde{p}_k,n)\; \right]\;.
\end{align}
This result can be derived from Eq.~\eqref{eq:h_operator_cs} by using 
the known expressions for the breaking of the Gribov-Lipatov relation 
at NLO QCD, see Sec.~6.4 of~\cite{Curci:1980uw}.
All remaining components of the subtraction formulae can be found in 
Appendix~C of~\cite{Catani:1996vz}.

\bibliography{journal}
\end{document}